\documentclass[preprint,showpacs,preprintnumbers,amsmath,amssymb]{revtex4}\newcommand{\figwidth}{0.5\textwidth}

\usepackage{epsfig}
\usepackage{graphicx}
\usepackage{dcolumn}
\usepackage{bm}

\renewcommand{\i}{\infty}   
\newcommand{\E}[1]{Eq.~(\ref{#1})}
\newcommand{\F}[1]{Fig.~\ref{#1}}
\renewcommand{\t}{\tau}        
\newcommand{\w}{\omega}  
\newcommand{\ba}{\begin{array}}
\newcommand{\ea}{\end{array}}
\newcommand{\g}{\gamma}  
\renewcommand{\o}{\Omega}  
\newcommand{\ra}{\rangle}
\newcommand{\la}{\langle}

\newcommand{\tlock}{\langle T_{\rm lock}\rangle}

\newcommand{\drm}{{\rm d}}
\def\mean#1{\langle#1\rangle}
\def\sigmastat{\sigma^{\ast}}
\def\omegaoutstat{\omega^{\ast}_{\rm out}}

\begin{document}


\title{Frequency and Phase Synchronization in Stochastic Systems}

\author{Jan A. Freund}
\email{freund@physik.hu-berlin.de}
\author{Lutz Schimansky-Geier}
\affiliation{
  Institut f\"ur Physik, Humboldt-Universit\"at zu Berlin,
  Invalidenstr.~110, D-10115 Berlin, Germany
}

\author{Peter H\"anggi}
\affiliation{
  Institut f\"ur Physik, Universit\"at Augsburg,
  Universit\"atsstr.~1, D-86135 Augsburg, Germany
}

\date{\today}

\begin{abstract}
  The phenomenon of frequency and phase synchronization in stochastic
  systems requires a revision of concepts originally phrased in the
  context of purely deterministic systems. Various definitions of an
  instantaneous phase are presented and compared with each other with
  special attention payed to their robustness with respect to noise.
  We review the results of an analytic approach describing
  noise-induced phase synchronization in a thermal two-state system.
  In this context exact expressions for the mean frequency and the
  phase diffusivity are obtained that together determine the average
  length of locking episodes.
  A recently proposed method to quantify frequency synchronization in
  noisy potential systems is presented and exemplified by applying it
  to the periodically driven noisy harmonic oscillator.  Since this
  method is based on a threshold crossing rate pioneered by S.O. Rice
  the related phase velocity is termed Rice frequency.
  Finally, we discuss the relation between the phenomenon of
  stochastic resonance and noise-enhanced phase coherence by applying
  the developed concepts to the periodically driven bistable Kramers
  oscillator.
\end{abstract}
\maketitle
%
%
{\bf Studying synchronization phenomena in stochastic systems
  necessitates a revision of concepts originally developed for
  deterministic dynamics. This statement becomes obvious when
  considering the famous phase-locking effect \cite{arnold,Strato67}:
  unbounded fluctuations that occur, for instance, in Gaussian noise
  will always prevent the existence of a strict bound for the
  asymptotic phase difference of two systems. Nevertheless, a
  reformulation of the synchronization phenomenon in the presence of
  noise is possible by quantifying the average duration $\tlock$ of
  locking epochs that are disrupted by phase slips. In case that
  $\tlock\gg T_0$, where $T_0$ is some characteristic time of the
  dynamics, e.g.~the period of an external drive or the inverse
  of some intrinsic natural frequency, it is justified to speak about
  effective phase synchronization.}
\section{Introduction}
\label{sec:0}%
From the conceptual point of view different degrees of synchronization
can be distinguished: complete synchronization \cite{cs}, generalized
synchronization \cite{gs}, lag synchronization \cite{ls}, phase
synchronization \cite{ps1,ps2}, and burst (or train) synchronization
\cite{bs}. In the following, we focus our attention on phase
synchronization in stochastic systems that has attracted recent
interest for the following reason: in many practical applications the
dynamics of a system, though not perfectly periodic, can still be
understood as the manifestation of a stochastically modulated limit
cycle \cite{book,biophys}. As examples, we mention neuronal activity
\cite{neurons}, the cardiorespiratory system \cite{cardresp}, or
population dynamics \cite{popdynamics}.

Given a data set or some model dynamics there exists a variety of
methods to define an instantaneous phase $\phi(t)$ of a signal or a
dynamics. For a clear cut separation of deterministic and
noise-induced effects it is essential to assess the robustness of each
of these different phase definitions with respect to noise. Section
\ref{sec:1} is devoted to this issue. Since we do not distinguish
between dynamical and measurement noise \cite{dynmeasnoise} our
treatment is also tied to the question how synchronization can be
detected within any realistic experimental data.

The synchronization properties of a noisy system can be classified in
a hierarchical manner: stochastic phase locking always implies
frequency locking while the converse is not true in general. On the
other hand, small phase diffusivity is necessary but not sufficient
for phase synchronization. This will become clear in Sec.~\ref{sec:2}
when we review an analytic approach \cite{FreuNeiLsg2000} to
stochastic phase synchronization developed for a thermal two-state
system with transitions described by noise-controlled rates.

A recently proposed method \cite{Callenbach} to measure the average
phase velocity or frequency in stochastic oscillatory systems based on
Rice's rate formula for threshold crossings \cite{Rice44,Rice54} will
be presented and discussed in Sec.~\ref{sec:3}. The Rice frequency
proves to be useful especially in underdamped situations whereas the
overdamped limit yields only finite values for coloured noise. Its
relation to the frequency based on the widely used Hilbert phase
(cf.~Sec.~\ref{sec:1:phiH}) is discussed and illustrated.

In the final Sec.~\ref{sec:4} we connect the topic of stochastic
resonance (SR) \cite{SR_rev1,SR_rev2} with results on noise-enhanced
phase coherence \cite{nei_lsg98}. To this end we study the
synchronization properties of the bistable Kramers oscillator driven
externally by a periodic signal \cite{jung}. As a complement to the
frequently investigated overdamped limit, we consider here the
underdamped case employing the methods presented in Sec.~\ref{sec:3}.
\section{Phase definitions in the presence of noise}
\label{sec:1}
\subsection{Natural phase $\phi^N$}
\label{sec:1:phiN}
A phase occurs in a quite natural way when describing the cyclic
motion of an oscillator in phase space. Self-sustained oscillators are
nonlinear systems that asymptotically move on a limit cycle. The
instantaneous position in phase space can be represented through
instantaneous amplitude $a^N(t)$ and phase $\phi^N(t)$. A systematic
approach to relate the amplitude and phase dynamics to the dynamics
formulated in original phase space was developed by Bogoliubov and
Mitropolski \cite{BogMit61}. Their method starts from the following
decomposition of the dynamics
\begin{eqnarray}
  \label{eq:bogmist1a}
  \dot x\;&=&\;v\,,\\
  \label{eq:bogmist1b}
  \dot v\;&=&\,-\omega_0^2\,x + f(x,v,t,\xi,\ldots)
\end{eqnarray}
where the function $f$ comprises all terms of higher than first order
in $x$ (nonlinearities), velocity dependent terms (friction), and
noise. In their work Bogoliubov and Mitropolski considered the
function $f$ to be a small perturbation of order $\epsilon$ which
means that the system is weakly nonlinear and the noise or the
external forces are comparatively small as not to distort the harmonic
signal too much. The definition of an instantaneous phase proceeds by
expressing the position $x$ and the velocity $v$ in polar coordinates
$a^N$ and $\phi^N$
\begin{eqnarray}
  \label{eq:bogmist2a}
  x(t) &=& a^N(t)\cos\left[\phi^N(t)\right]\,,\\
  \label{eq:bogmist2b}
  v(t) &=& -\omega_0\; a^N(t) \sin\left[\phi^N(t)\right]
\end{eqnarray}
which yields by inversion \cite{inversion}
\begin{eqnarray}
  \label{eq:bogmist3a}
  a^N(t) &=& \sqrt{ x^2(t)+\left [ v(t)/\w_0 \right ]^2 }\,,\\
  \label{eq:bogmist3b}
  \phi^N(t) &=& \arctan \left[ -\frac{v(t)/\w_0}{x(t)} \right] .
\end{eqnarray}
It should be noted that a meaningful clockwise rotation in the
$x,v$-plane determines angles to be measured in a specific way
depending on the sign of $\omega_0$. Using Eqs.~(\ref{eq:bogmist2a}),
(\ref{eq:bogmist2b}), (\ref{eq:bogmist3a}) and (\ref{eq:bogmist3b}) it
is straightforward to transform the dynamics in $x$ and $v$,
Eqs.~(\ref{eq:bogmist1a}) and (\ref{eq:bogmist1b}), into the following
dynamics for $a^N$ and $\phi^N$ \cite{Strato67,HangRise83}
\begin{eqnarray}
  \label{eq:rdot}
  \dot a^N &=& 
  - \frac{f\left( a^N\cos(\phi^N),-\omega_0\, a^N\sin(\phi^N),t,\xi\right)}
  {\omega_0}\sin(\phi^N)\\
  \label{eq:phidot}
  \dot\phi^N &=& \omega_0
  - \frac{f\left( a^N\cos(\phi^N),-\omega_0\, a^N\sin(\phi^N),t,\xi\right)}
  {\omega_0 a^N}\cos(\phi^N).
\end{eqnarray}
The line $x=0$ corresponds to angles $\phi^N=\pi/2 + n\pi, n\in N$. As
can be read off from Eq.~(\ref{eq:phidot}), the phase velocity always
assumes a specific value for $x=0$ \cite{divf}, i.e.,
\begin{equation}
  \dot\phi^N(x=0) = \omega_0\,.
\end{equation}
This has the following remarkable consequence. We see that even in the
presence of noise passages through zero in the upper half plane $v>0$
are only possible from $x<0$ to $x>0$, in the lower half plane only
from $x>0$ to $x<0$. This insight becomes even more obvious from a
geometrical interpretation: as the noise exclusively acts on the
velocity $v$, cf.~Eq.~(\ref{eq:bogmist1b}), it can only effect changes
in the vertical direction (in $x,v$-space). Along the vertical line
$x=0$, however, the angular motion possesses no vertical component
while radial motion is solely in the vertical direction and,
therefore, only affected by the noise. From this we conclude that
between subsequent zero crossings of the coordinate with positive
velocity the phase has increased by an amount of $2\pi$. This finding
establishes a simple operational instruction how to measure the
average phase velocity of stochastic systems. We will come back to
this point in Sec.~\ref{sec:3}

\subsection{Linear interpolating phase $\phi^L$}
\label{sec:1:phiL}
As we have just seen zero crossings can be utilized to mark the
completion of a cycle. This can be generalized to the crossings of an
arbitrary threshold with positive velocity or even to the crossing of
some separatrix. In this connection the concept of isochrones of a
limit cycle has to be mentioned \cite{isochrones}. All of these
extensions of the natural phase require a thorough knowledge of the
dynamics and the phase space structure. In many practical applications,
however, the detailed phase portrait is not known. Instead, one is
given a data series exhibiting a repetition of characteristic marker
events, e.g.~the spiky peaks of neural activity, the R-peaks
of an electrocardiogram, or pronounced maxima as found in population
dynamics. These marker events can be used to pinpoint the completion
of a cycle, $k$, and the beginning of a subsequent one, $k+1$. It is
then possible to define an instantaneous phase $\phi^L(t)$ by linear
interpolation, i.e.,
\begin{equation}
  \phi^L(t)= \frac{t-t_k}{t_{k+1}-t_k}2\pi \;+\;k\,2\pi
  \qquad (t_k\le t<t_{k+1})\,
  \label{linphase}
\end{equation}
where the times $t_k$ are fixed by the marker events.
Reexpressing the time series $x(t)$ of the system as
\begin{equation}
  x(t)=a^L(t)\,\cos[\phi^L(t)],
\end{equation}
then defines an instantaneous amplitude $a^L(t)$. The benefit of such
a treatment is to reveal a synchronization of two or more such
signals: whereas the instantaneous amplitudes and, therefore, the time
series might look rather different, the phase evolution can display
quite some similarity. If the average growth rates of phases match
(notwithstanding the fact that phases may diffuse rapidly) the result
is termed frequency locking. Small phase diffusion, in addition to
frequency locking, means that phases are practically locked during
long episodes that occasionally are disrupted by phase slips caused by
sufficiently large fluctuations. This elucidates the meaning of
effective phase synchronization in stochastic systems.

As should be clear from its definition the linear phase relies on the
clear identification of marker events. With increasing noise intensity
this identification will fail since sufficiently large fluctuations
may either mask true or imitate spurious marker events. On the other
hand, in some cases, e.g.~for excitable systems, fluctuations can be
essential for the generation of marker events, i.e., marker events may
be noise-induced.

As a final remark, let us mention that relative maxima of a
differentiable signal $x(t)$ correspond to positive-going zeros of its
derivative $\dot x(t)$. However, this seemingly trivial connection is
overshadowed by complications if the derivative itself is a non-smooth
function that does not allow to easily extract the number of zero
crossings (cf.~Sec.~\ref{sec:3} and Fig.~\ref{f_oscillator}).

\subsection{Hilbert phase $\phi^H$}
\label{sec:1:phiH}
In situations where a measured signal $x(t)$ exhibits a lot of
irregularity it is not quite clear how to define a phase -- the signal
might look far from a perturbed harmonic or even periodic one and
marker events cannot be identified unambiguously. The concept of the
analytic signal as introduced by Gabor \cite{AnaSig} offers a way to
relate the signal $x(t)$ to an instantaneous amplitude $a^H(t)$ and a
phase $\phi^H(t)$. The physical relevance of a such constructed phase
is a question of its own; for narrow-band signals or harmonic noise it
has a clear physical meaning whereas the general case requires further
considerations (cf.~Appendix A2 in \cite{survey}).

The analytic signal approach extends the real signal $x(t)$ to a
complex one $z(t)=x(t)+i\,y(t)=a^H(t)\exp[\phi^H(t)]$ with the
imaginary part $y(t)$ resulting from an appropriate transformation of
the real part $x(t)$. Instead of taking $y(t) = -\dot x(t)/\w_0$ as
for the natural phase we search $y(t)$ as the result of a convolution
of $x(t)$ with some appropriate kernel $K(t)$, i.e.,
\begin{equation}
  y(t)=\int\limits_{-\infty}^{\infty} x(\tau)K(t-\tau)\; d\tau\,.
 \label{eq:convol} 
\end{equation}
Now, {\em appropriate} means that the kernel has to be chosen such
that the method reproduces the phase of a harmonic signal. Applying
the convolution theorem it is easy to see that the Fourier transform
$Y(\omega)$ of the transformed signal $y(t)$ should be related to the
Fourier transform $X(\omega)$ of the original signal $x(t)$ by a phase
shift that transforms a cosine into a sine, i.e.,
\begin{equation}
  \label{eq:fourhilb}
  Y(\omega)= -i \mbox{sgn}(\omega) X(\omega)\,.
\end{equation}
where sgn($\cdot$) is the sign function. By an inverse Fourier
transform we thus find that $K(t)=1/(\pi t)$ which implies that $y(t)$
is related to $x(t)$ via the Hilbert transform
\begin{equation}
  y(t)=x^H(t)=\frac{1}{\pi}\;
  P\!\!\int\limits_{-\infty}^{\infty} \frac{x(\tau)}{t-\tau}\; d\tau\,.
  \label{eq:hilberttrafo}
\end{equation}
The symbol $P$ in front of the integral in Eq.~(\ref{eq:hilberttrafo})
is a reminder that the integral has to be evaluated in the sense of
the Cauchy principal value. The fact that the Hilbert phase
\cite{inversion}
\begin{equation}
  \label{eq:hilphase}
  \phi^H(t)=\arctan\left[\frac{x^H(t)}{x(t)}\right]
\end{equation}
arises as the result of a convolution instead of a differentiation
makes it less sensitive to short-lived small fluctuations. This
observation was already reported by Vainstein and Vakman
\cite{VainVak83}.

Moreover, the construction by a convolution brings the Hilbert phase
in close contact with the wavelet transform that is widely used to
compute a time-dependent spectral decomposition of non-stationary
signals \cite{wavelet}. By virtue of Eq.~(\ref{eq:fourhilb}) it is
evident that the Fourier transform of the analytic signal $Z(\omega)$
satisfies the relation
\begin{equation}
  \label{eq:anasigfourier}
  Z(\omega)=\left[1+\mbox{sgn}(\omega)\right]\,X(\omega)
  =2\,\theta(\omega)\,X(\omega)
\end{equation}
where $\theta(\cdot)$ is the step function. Equation
(\ref{eq:anasigfourier}) shows that all positive Fourier components of
the original signal contribute with equal weights. Selecting, instead,
a certain frequency band by using a subsequent Gaussian filter (in
frequency space) corresponds to a convolution with the Morlet wavelet
(Gabor function). Generalized phase definitions and the wavelet
transform were employed to detect phase-synchronous activity in the
brain \cite{neusync} and in a chaotic laser array \cite{lassync}.

\subsection{Discrete phase $\phi^D$}
\label{sec:1:phiD}
\begin{figure}
  \centerline{\epsfig{figure=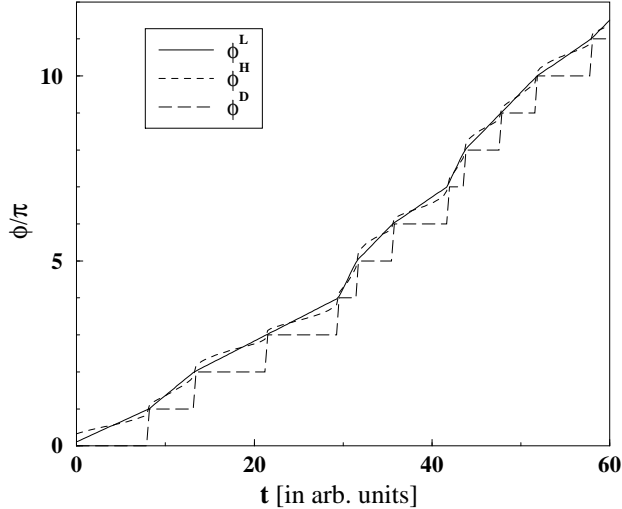,width=\figwidth}}
  \caption{The linear interpolating phase, the Hilbert phase and
    the discrete phase agree in the description of a dichotomous
    switching process}
  \label{fig:3phases}
\end{figure}

Multistability is one of the crucial consequences of nonlinearity and
plays a dominant role for many important topics, e.g.~evolution,
information processing and communication, pattern formation, etc.
Frequently, fluctuations play a benificial role in that they effect
transitions between the different states that are directly tied to the
performance of a task. The phenomenon of SR \cite{SR_rev1,SR_rev2},
for instance, can be observed in a bistable system.  Typically, the
information that is processed in a bistable system does not require to
keep track of a continuum but is rather contained in the switching
events between the two states \cite{McNamWies89}.  Hence, it is
desirable to link the dichotomous switching process to a description
in terms of an instantaneous phase.  Since switching to and fro
constitutes one cycle and, thus, corresponds to a phase increment of
$2\pi$ the linear interpolating phase $\phi^L$ can be readily
constructed employing switchings as marker events of half-cycles.
Furthermore, the Hilbert phase can be easily computed. Alternatively,
it is possible to use the switching process (between $-1$ and $+1$)
and construct a discrete instantaneous phase $\phi^D(t)$ changing
discontinuously at the switching events.  The last-mentioned
instantaneous phase is obtained simply by multiplying the number of
switches (or the number of renewals in renewal theory \cite{Cox67})
$k(t)$ by the value $\pi$, i.e., $\phi^D(t)=k(t)\pi$.  The
instantaneous state, in turn, can be obtained from the discrete phase
via $x(t)=\exp[i\phi^D(t)]$.  In Fig.~\ref{fig:3phases} we show how
the three alternative phases $\phi^L,\phi^H$ and $\phi^D$ agree in the
description of a dichotomous switching process. Note that the natural
phase $\phi^N(t)$ is related to the underlying process in real phase
space and, hence, cannot be deduced from the two-state signal.
The advantage of the discrete phase is that it allows an
analytic treatment of effective phase synchronization in stochastic
bistable systems. This will be addressed in Sec.~\ref{sec:2}.

The robustness of the discrete phase with respect to noise comes into
play not when making the ``transition'' from the switching process to
the instantaneous phase $\phi^D(t)$ but when constructing the
switching events from the continuous stochastic trajectory. At the
level of a Markovian switching process noise enters only via its
intensity that changes transition rates.

\subsection{Phases in higher dimensional systems}
\label{sec:1:d>1}
In systems of dimension $d>1$ there are various ways how to define one
or even more instantaneous phases: projecting the $2d$-dimensional
phase space onto 2-dimensional surfaces, choosing Poincar\'{e}
sections or computing the Hilbert phase for each of the coordinates.
Many of the methods can only be done numerically and always require to
consider the dynamics in detail in order to check whether a made
choice is appropriate. We will not elaborate these details here but
refer to \cite{survey} (especially Chap.~10) and references therein.

\section{Analytic treatment of a driven noisy bistable system}
\label{sec:2}
\subsection{Setup of the doubly dichotomous system}
\label{sec:2:setup2x2}
In this section we consider a stochastic bistable system, for instance
a noisy Schmitt trigger, which is driven externally either by a
dichotomous periodic process (DPP) or a dichotomous Markovian process
(DMP). The dichotomous character of the input shall be either due to a
two-state filtering or be rooted in the generation mechanism of the
signal. The bistable system generates a dichotomous output signal. For
convenience we choose to label input and output states with values
$+1$ and $-1$ respectively. The DPP is completely specified by its
angular frequency $\Omega=\pi/T_0$ where $T_0$ is the half-period.
Accordingly, the DMP is fully characterized by its average switching
rate $\gamma=T_0^{-1}$.

Since both input and output are two-state variables it is possible to
study phase synchronization in terms of discrete \footnote{for reasons
  of convenience we will suppress the index $D$ indicating the
  discrete phase considered throughout this section}
input and output phases $\phi_{\rm in}(t)$ and $\phi_{\rm out}(t)$
respectively. Consequently, also the phase difference
$\phi(t)=\phi_{\rm out}(t)-\phi_{\rm in}(t)$ is a discrete quantity
that can assume positive and negative multiples of $\pi$. From the
definition of the phase difference $\phi$ it follows that each
transition between the output states {\em increases} $\phi$ by $\pi$
whereas each transition between the input states {\em reduces} $\phi$
by $\pi$.
 
Transitions between the input states are governed by the rates
\begin{equation}
  \label{inputrates}
  W^{DMP}_{\pm}=\gamma\quad\mbox{and}\quad
  W^{DPP}_{\pm}(t)=\sum\limits_{n=-\infty}^{\infty} \delta(t-t_n)
\end{equation}
where a single realization of the DPP is characterized by
deterministic switching times $t_n=(n\pi+\varphi_0)\Omega^{-1}$.
Here, $\varphi_0$ is the initial phase of the input signal rendering
the periodic process non-stationary (cyclo-stationary). To achieve
strict stationarity we average the periodic dynamics with respect to
$\varphi_0$ which is equidistributed over the interval $[0,2\pi)$.

In the absence of an input signal the two states are supposed to be
symmetric and the hopping rates for both directions are identical and
completely determined by a prefactor $\alpha_0$, the energy barrier
$\Delta U$, and the noise intensity $D$. The central assumption of
our analysis is that the input signal modifies the transition rates
of the output solely through the phase difference $\phi$ in the
following way
\begin{equation}
  \label{kram1}
  W^{out}(\phi)= g(\phi) = \alpha_0 \,\exp\left[-\frac{\Delta
      U + A \, \sigma(\phi)}{D}\right]\,,
\end{equation}
where the function $\sigma(\phi)=\cos(\phi)=\pm 1$ and the amplitude
$A <\Delta U=0.25$ to keep the signal subthreshold. This definition
introduces two noise-dependent time scales
\begin{equation}
  \label{a1a2}
  a_1=\alpha(D) \, \exp\left(-\frac{A}{D}\right)
  \quad\mbox{and}\quad
  a_2=\alpha(D) \, \exp\left(\frac{A}{D}\right)
\end{equation}
with $\alpha(D) = \alpha_0 \exp(-\Delta U/D)$. The function
$\sigma(\phi)$ favours phase differences with even multiples of $\pi$,
i.e., in-phase configurations.

A description of the stochastic evolution of the phase difference is
based on the probabilities $P(\phi,t|\phi_0,t_0)$ to experience a
phase difference $\phi$ at time $t$ conditioned by a phase difference
$\phi_0$ at time $t_0$. Due to the discrete character of $\phi$
(allowing only for multiples of $\pi$) we briefly denote $P_{\rm
  k}=P(\phi=k\pi,t|\phi_0,t_0)$. Then the probabilistic evolution
operator reads with $g_{\rm k}=g( \phi=k\pi)$ from (\ref{kram1})
\begin{equation}
  \label{ph1}
  \frac{\partial P_{\rm k}(t)}{\partial t}= 
  \hat{L} P_{\rm k}(t) + g_{\rm k-1} P_{\rm k-1}(t) - 
  g_{\rm k} P_{\rm k}(t)\,.
\end{equation}
While the last two terms on the right hand side of Eq.~(\ref{ph1})
account for the change of $\phi$ due to transitions of the output the
operator $\hat{L}$ reflects switches of the input
\begin{equation}
  \label{L}
  \hat{L}P_{\rm k} = W_{\pm}\,(P_{\rm k+1}-P_{\rm k})
\end{equation}
with the related input switching rates $W_{\pm}$ given by
Eq.~(\ref{inputrates}). As mentioned above the non-stationary
(cyclo-stationary) character of the DPP can be cured by averaging over
the initial phase $\varphi_0$. Since ``temporal'' and ``spatial''
contributions in Eq.~(\ref{L}) are separable we can perform this
average prior to the calculation of any moment of $\phi$ yielding
\begin{equation}
  \label{average} 
  \langle W^{DPP}_{\pm}\rangle_{\varphi_0} =
  \int\limits_0^{2\pi}\frac{\drm\varphi_0}{2\pi}
  \sum\limits_{n=-\infty}^{\infty}
  \delta\left(t-\frac{n\pi+\varphi_0}{\Omega}\right) = 
  \frac{\Omega}{\pi}\,.
\end{equation}
From Eq.~(\ref{average}) we see that the $\varphi_0$-averaged DPP
formally looks equivalent to a DMP with the transition rate
$\Omega/\pi$. Of course, initial phase averaging does not really turn
a DPP into a DMP. The subtle difference is that while $\phi$-moments
of the DMP continuously change in time related $\phi$-averages of the
DPP (before the $\varphi_0$-average) are still discontinuous, hence,
temporal derivatives of functions of $\mean{\phi}$ have to be computed
with care before initial phase averaging
\cite{FreuNeiLsg2000,FreuNeiLsg2001}.

\subsection{Noise-induced frequency locking}
\label{sec:2:nifl}
Using standard techniques \cite{vankampen} we can derive the evolution
equation for the mean phase difference $\mean{\phi}$ from
Eq.~(\ref{ph1})
\begin{eqnarray}
  \label{ph2b}
  \mean{\dot{\phi}} &=& 
  -\mean{\omega_{\rm in}} + \mean{\omega_{\rm out}}\\
  &=&
  -\mean{\omega_{\rm in}}
  + \frac{\pi}{2} (a_1 + a_2) 
  - \frac{\pi}{2} (a_2 - a_1) \mean{\sigma}\,.  
\end{eqnarray}
Here, $\mean{\omega_{\rm in}}$ denotes the average frequency of the
input phase and equals $\gamma\pi$ for the DMP and $\Omega$ for the
DPP. Assuming higher moments uncoupled, i.e., $\mean{\sigma(\phi)}
\propto \sigma(\mean{\phi})$, Eq.~(\ref{ph2b}) is Adler's equation
\cite{adler} arising in the context of phase locking. Note that here
both the frequency mismatch $\Delta =-\mean{\omega_{\rm
    in}}+\frac{\pi}{2}(a_1 + a_2)$ and the synchronization bandwidth
$\Delta_s=\frac{\pi}{2}(a_2 - a_1)$ are noise dependent. This
elucidates the opportunity to achieve {\em noise-induced} frequency
and effective phase locking. For the short-time evolution a necessary
condition for locking is $|\Delta|<\Delta_s$ which defines ``Arnold
tongues'' \cite{arnold} of synchronization in the $A$ vs.~$D$ plane.

The kinetic equation for $\mean{\sigma}$ can be evaluated explicitly
yielding
\begin{equation}
  \label{sigdot}
  \mean{\dot{\sigma}} = 
  -\left[2\frac{\mean{\omega_{\rm in}}}{\pi} + a_1+a_2\right]
  \mean{\sigma} + a_2-a_1\,.
\end{equation}
From Eq.~(\ref{sigdot}) we see that $\mean{\sigma}$ approaches a
stationary value
\begin{equation}
  \label{ph4}
  \mean{\sigmastat}=
  \frac{a_2-a_1}{2\displaystyle\frac{\mean{\omega_{\rm in}}}{\pi}
    +a_1+a_2} 
\end{equation}
that exactly coincides with the stationary correlation coefficient
between the input and output \cite{nei_lsg99}. The relaxation time is
given by $\tau=[2\mean{\omega_{\rm in}}/\pi+a_1+a_2]^{-1}$. Hence, the
stationary output phase velocity can be achieved from Eq.~(\ref{ph2b})
by insertion of Eq.~(\ref{ph4}) yielding
\begin{equation}
  \label{ph5}
  \mean{\omegaoutstat} = \frac{\pi}{2} (a_1+a_2) - 
  \frac{\pi}{2} (a_2-a_1) \mean{\sigmastat}\,.
\end{equation}
This expression is in agreement with similar results derived in the
context of resonant activation \cite{chris}.
\begin{figure}
  \centerline{\epsfig{figure=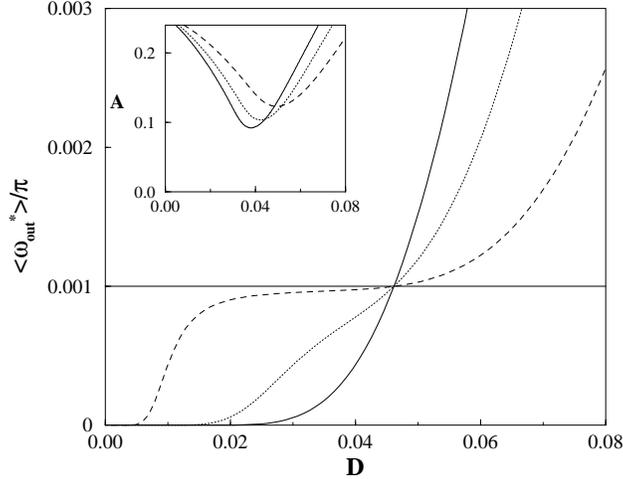,width=\figwidth}}
  \caption{Mean output switching frequency (in units of $\pi$)
    vs.~noise intensity for the DMP [or the DPP] signal for
    $\gamma=0.001$ [or $\Omega/\pi=0.001$ respectively] and for three
    amplitudes $A$: 0 (solid), 0.1 (dotted), 0.2 (dashed) ($\Delta
    U=0.25$, other parameters see text). ``Arnold tongues'', defined
    by the demand of ``sufficiently small slopes'' (see text), are
    shown in the inset for various $\gamma$: 0.001 (solid), 0.002
    (dotted), 0.005 (dashed).}
  \label{fig:epl1}
\end{figure}
In Fig.~\ref{fig:epl1} we depict the mean output switching rate
$\mean{\omegaoutstat}$ as a function of the noise intensity $D$ for
several values of the input signal amplitude $A$. With increasing
amplitude the region of frequency locking, i.e., the region of $D$ for
which $\mean{\omegaoutstat}\approx\mean{\omega_{\rm in}}$, widens
(cf.~also Fig.~9 in \cite{asyschmitt}). For most of these intensities
the bistable system possesses rates that do not obey the time scale
matching condition; nevertheless, on the average the output switching
events are entrained by the input signal. As mentioned before the
whole effect is non-linear: If detuning becomes too large the output
gets desynchronized and returns to its own dynamics.

Besides many alternatives, we defined the width of the frequency
locking region by requiring the slope of the curves to be less than
30\% of the slope for $A=0$ at the point where the output switching
rate coincides with $\gamma$ [or $\Omega/\pi$ respectively] --
simultaneously disqualifying the initial flat region for small $D$.
The resulting ``Arnold tongues'' (compatible with data from
\cite{shulgin,nei_lsg98,nei_lsg99}) are shown in the inset of
Fig.~\ref{fig:epl1}. It can be seen that frequency locking
necessitates to exceed a minimal amplitude that shifts to lower values
for slower signals \cite{shulgin}. Let us emphasize that the frequency
locking region groups around the noise intensity that satisfies the
time scale matching condition (in our case by definition $\omega_{\rm
  in} = \omegaoutstat(D)$); for a harmonic input this range of $D$
also maximizes the spectral power amplification, in contrast to values
of $D$ where the signal-to-noise ratio attains its maximum ($D \propto
\Delta U$) \cite{SR_rev1,SR_rev2}.

\subsection{Phase locking and effective synchronization}
\label{sec:2:nips}
\begin{figure}
  \centerline{\epsfig{figure=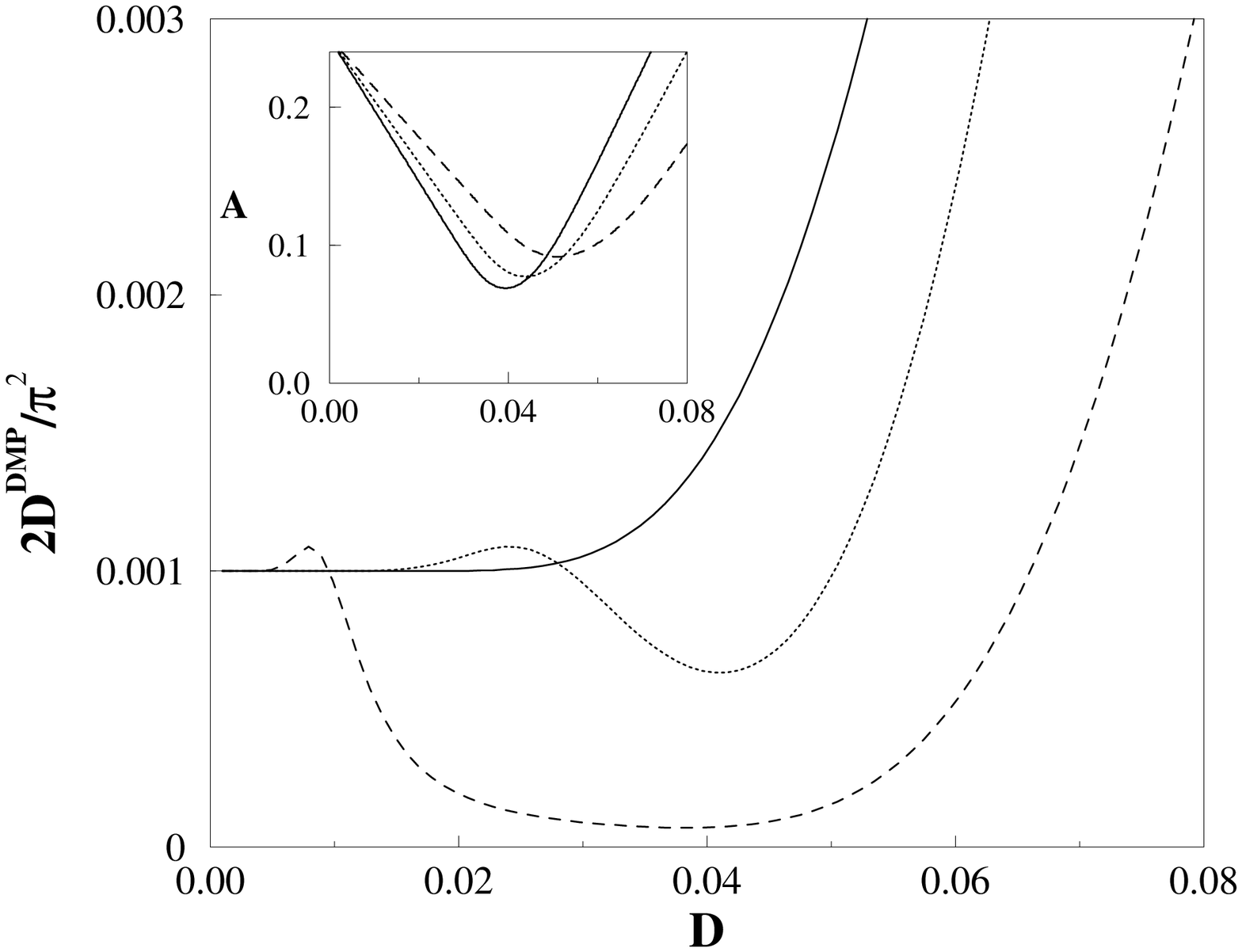,width=\figwidth}}
  \centerline{\epsfig{figure=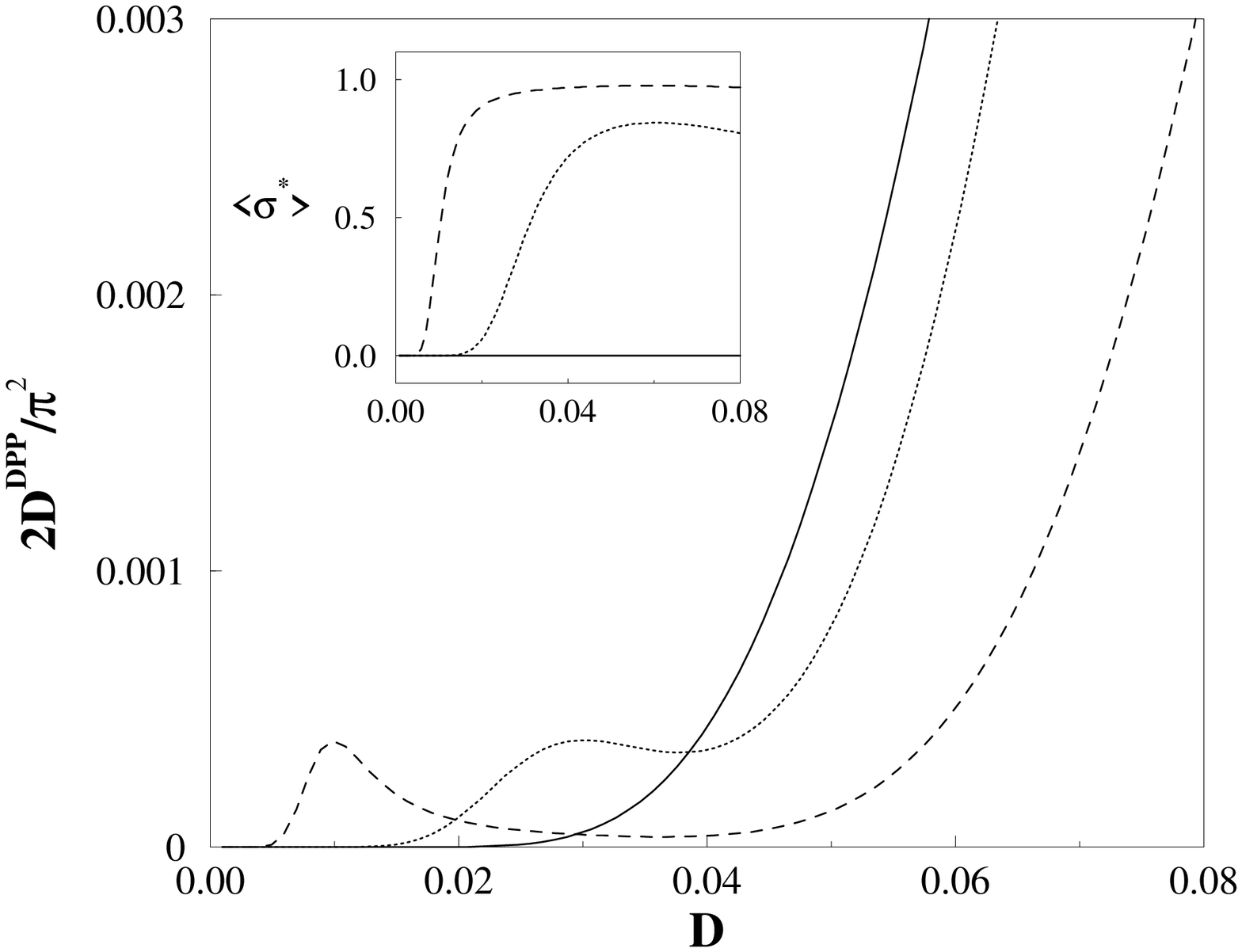,width=\figwidth}}
  \caption{Effective diffusion coefficients ${\cal D}^{DMP}$
    (top) and ${\cal D}^{DPP}$ (bottom) of the instantaneous phase
    difference $\phi$ as functions of the noise intensity $D$ for
    $\gamma=\frac{\Omega}{\pi}=0.001$ and for three amplitudes $A$: 0
    (solid), 0.1 (dotted), 0.2 (dashed) (other parameters as in
    Fig.~\ref{fig:epl1}). The values at $D=0$ are determined solely by
    the input diffusion and, hence, vanish for strictly periodic
    signals.  Defining the region of phase locking by the demand that
    ${\cal D}^{DMP}<{\cal D}^{DMP}_{\rm in}$ yields the tongues
    depicted in the upper inset which, as in Fig.~\ref{fig:epl1},
    reveals a critical amplitude varying with $\gamma$: 0.001 (solid),
    0.002 (dotted), 0.005 (dashed). As can be seen from the lower
    inset phase locking occurs for considerably large
    $\langle\sigma^{\ast}\rangle$ only (curves for amplitudes $A$: 0
    (solid), 0.1 (dotted), 0.2 (dashed)).}
  \label{fig:epl2}
\end{figure}

The phenomenon of phase locking can be demonstrated by considering the
diffusion coefficient ${\cal D}$ of the phase difference, achieved as
the time derivative of the variance $\frac{1}{2}\partial_{\rm t}
[\mean{\phi^2}-\mean{\phi}^2]$. Performing the calculation for both
the DMP and DPP yields
\begin{equation}
  {\cal D} = {\cal D}_{\rm in}+\frac{\pi^2}{2}
  \Bigg[
  \frac{\mean{\omega_{\rm out}}}{\pi}
  -(a_2-a_1)
  \frac{\mean{\phi\sigma}-\mean{\phi}\mean{\sigma}}{\pi}
  \Bigg]
  \label{d*p}
\end{equation}
with ${\cal D}_{\rm in}=\frac{\pi^2}{2}\gamma$ for the DMP and ${\cal
  D}_{\rm in}=0$ for the DPP. The stationary correlator
$\mean{(\delta\phi\delta\sigma)^{\ast}}$, i.e., the asymptotic
limit of $\mean{\phi\sigma}-\mean{\phi}\mean{\sigma}$, can be computed
from the corresponding kinetic equation \cite{FreuNeiLsg2000}.
Inserting $\mean{(\delta\phi\delta\sigma)^{\ast}}$ into
Eq.~(\ref{d*p}) we thus find for the DMP (cf.~Fig.~\ref{fig:epl2} top)
\begin{eqnarray}
  {\cal D}^{DMP} &=&
  \frac{\pi^2}{2}
  \Bigg[
  \gamma +
  \frac{\mean{\omegaoutstat}}{\pi}
  -\left(2\gamma-(a_1+a_2)\right)\mean{\sigmastat}^2\nonumber\\
  &&
  \hspace*{1.cm}
  -\frac{1}{2}(a_2-a_1)\mean{\sigmastat}(1+\mean{\sigmastat}^2)
  \Bigg]
  \label{dmp2}
\end{eqnarray}
and for the DPP (cf.~Fig.~\ref{fig:epl2} bottom)
\begin{eqnarray}
  {\cal D}^{DPP} &=&
  \frac{\pi^2}{2}
  \Bigg[
  \frac{\mean{\omegaoutstat}}{\pi}
  -\left(2\frac{\Omega}{\pi}-(a_1+a_2)\right)\mean{\sigmastat}^2\nonumber\\
  &&
  \hspace*{.1cm}
  -\frac{1}{2}(a_2-a_1)\mean{\sigmastat}(1+\mean{\sigmastat}^2)
  +\frac{\Omega}{\pi}\mean{\sigmastat}
  \Bigg]
  \label{dpp2}
\end{eqnarray}
with $\mean{\sigmastat}$ given by Eq.~(\ref{ph4}) and
$\mean{\omegaoutstat}$ by Eq.~(\ref{ph5}).
Both Eqs.~(\ref{dmp2}) and (\ref{dpp2}) possess the same structure
${\cal D} = {\cal D}_{\rm in} + {\cal D}_{\rm out} - {\cal D}_{\rm
  co}$ with ${\cal D}_{\rm out}=\frac{\pi}{2}\langle\omega^{\ast}_{\rm
  out}\rangle$. Since ${\cal D}_{\rm out}$ is never decreasing (with
increasing noise intensity) the same is true for the sum of the first
two terms. The possibility of synchronized input-output jumps is
rooted in ${\cal D}_{\rm co}$. Since this term comprises only
contributions scaling with powers of $\langle\sigma^{\ast}\rangle$,
which itself rapidly vanishes for small $D$ (cf.~inset of
Fig.~\ref{fig:epl2} bottom), we first observe an increase of ${\cal
  D}$. An increase of $\langle\sigma^{\ast}\rangle$ signals the
coherent behaviour of input and output and, consequently, endows
${\cal D}_{\rm co}$ with considerable weight to outbalance the
increase of ${\cal D}_{\rm out}$. As can be seen from the inset of
Fig.~\ref{fig:epl2} bottom a negative slope is initiated only for a
sufficiently large $\langle\sigma^{\ast}\rangle$.  However, the range
of high input-output correlation $\langle\sigma^{\ast}\rangle$ does
not determine the range of low diffusion coefficients since at rather
high noise intensities $D$ the output switches with a large variance
and thus, finally dominates over the ordering effect of ${\cal D}_{\rm
  co}$.

Plotting the boundaries of the region where ${\cal D}^{DMP}<{\cal
  D}^{DMP}_{\rm in}$ defines the tongues depicted in the inset of
Fig.~\ref{fig:epl2} top. As for the ``Arnold tongues'' in
Fig.~\ref{fig:epl1} a minimal amplitude varies with the mean input
switching rate $\mean{\omega_{\rm in}}$ and shifts to lower values
when considering slower signals. It is worth mentioning that the
addition of an independent dichotomous noise that modulates the
barrier $\Delta U$ can drastically reduce this minimal amplitude if
this second dichotomous noise switches faster than the external signal
\cite{robrozen}.
  
\begin{figure}
  \centerline{\epsfig{figure=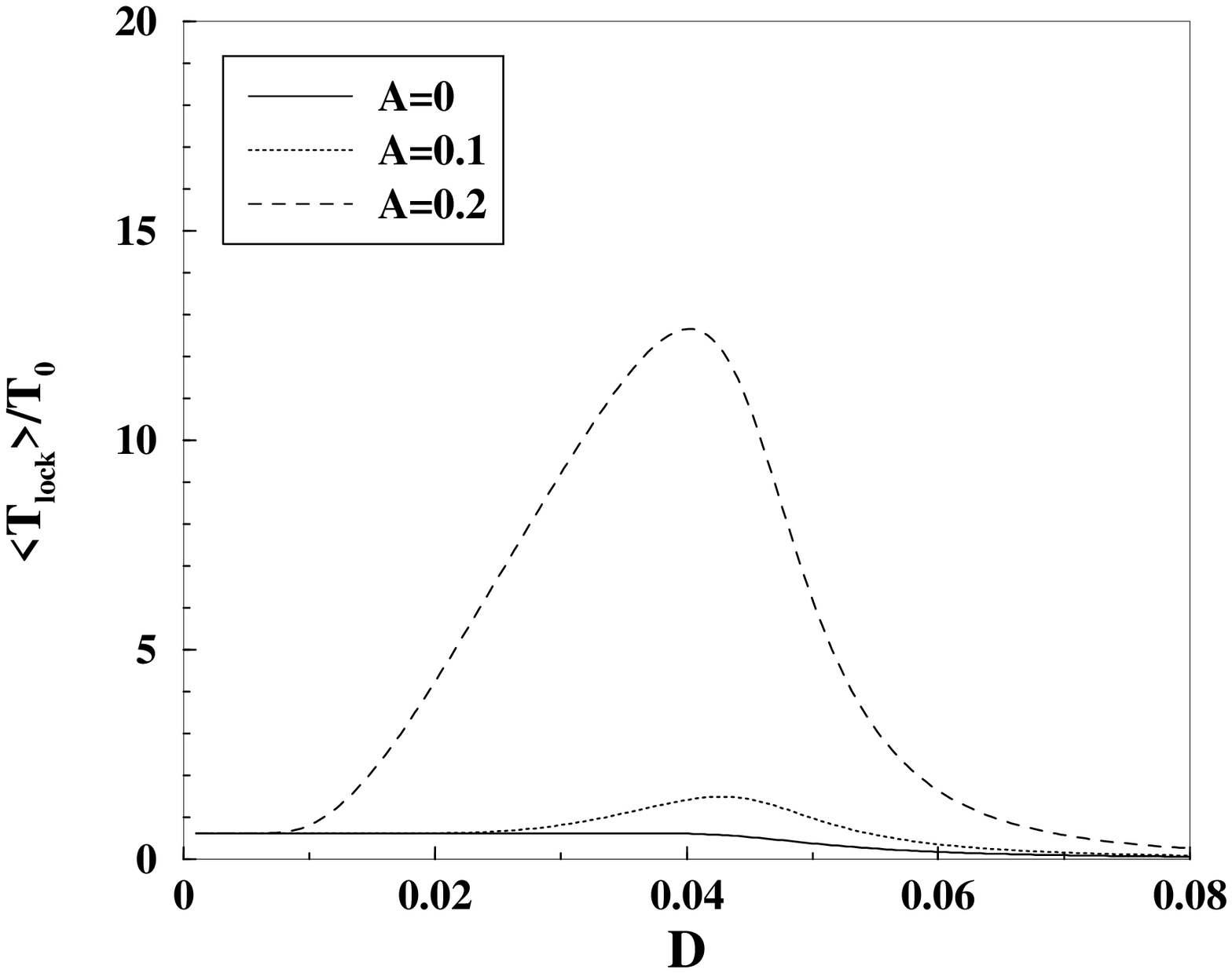,width=\figwidth}}
  \centerline{\epsfig{figure=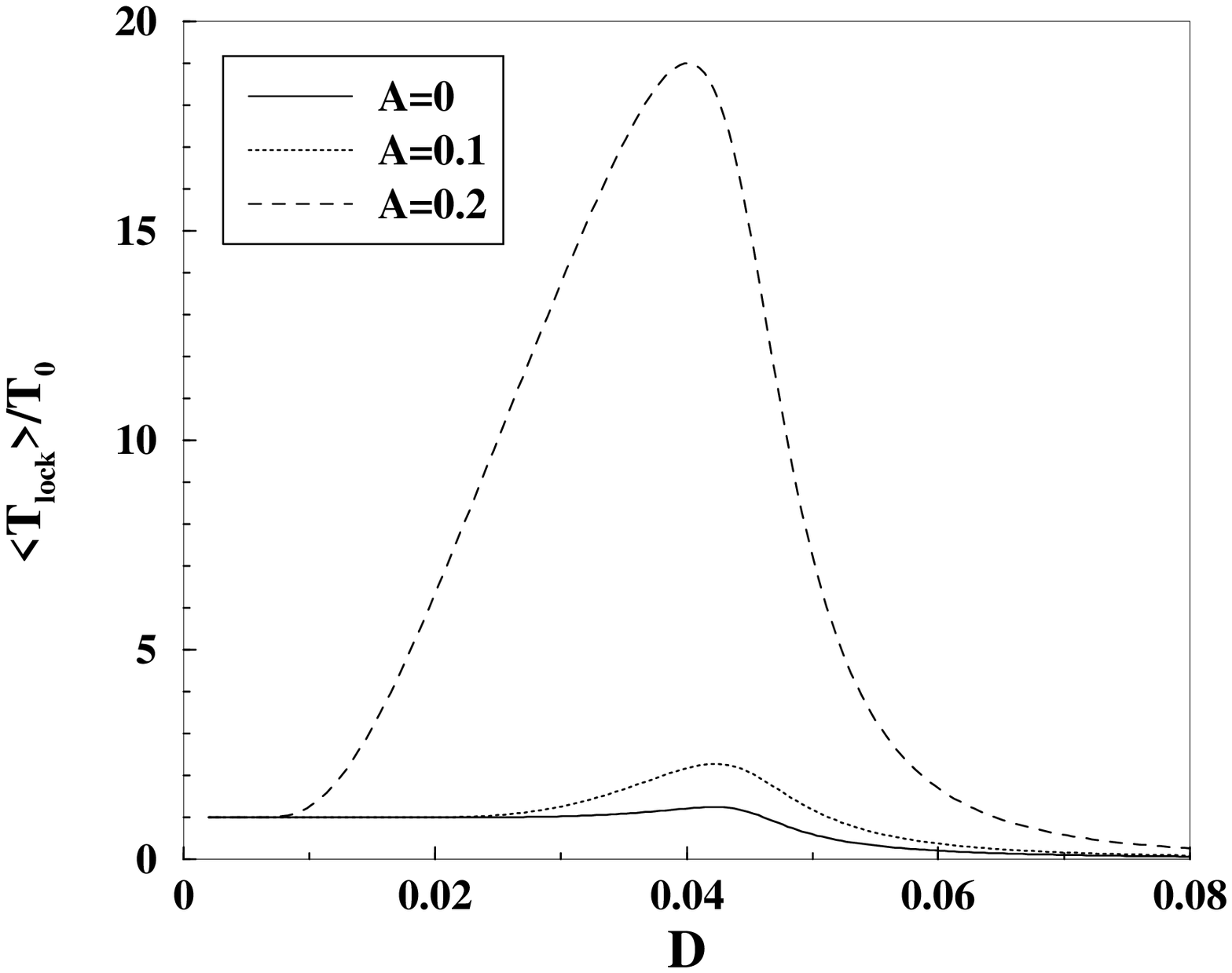,width=\figwidth}}
  \caption{The normalized average duration of locking episodes $\tlock$ 
    (cf.~Eq.~(\ref{eq:tlock})) for the DMP (top) and the DPP (bottom)
    exhibits an enormous maximum for intermediate values of noise intensity
    thus proving noise-induced phase synchronization}
  \label{fig:tlock}
\end{figure}

\begin{figure}
  \centerline{\epsfig{figure=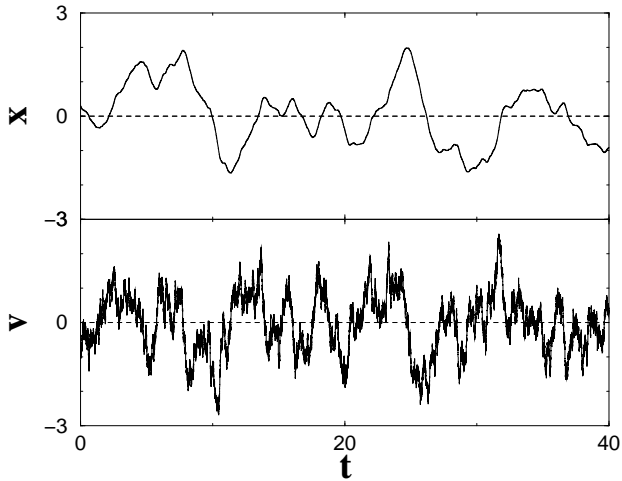,angle=-90,width=\figwidth}}
  \caption{Position $x$ and velocity $v$ of the undriven noisy harmonic
    oscillator \E{e_harmonicosc} with friction coefficient $\gamma=1$,
    and natural frequency $\w_0=1$. Whereas the position $x$ is smooth
    the velocity $v$ is continuous but nowhere differentiable.
    Counting of zero crossings is, consequently, only possible for the
    $x$-coordinate.}
    \label{f_oscillator}
\end{figure}

The minimum of ${\cal D}$ observed in the region of frequency locking
can be equivalently expressed as a pronounced maximum of the average
duration of locking episodes $\tlock$. To show this we note that a
locking episode is ended by a phase slip whenever the phase difference
has changed, i.e., increased or decreased, by the order of $\pi$, or
\begin{equation}
  \label{eq:tlock}
  \mean{\phi^2}=\mean{\dot\phi}^2\tlock^2 + 2{\cal D}\tlock = \pi^2\,.
\end{equation}
This quadratic equation can be solved for $\tlock$
\cite{FreuNeiLsg2001} and by inserting the noise-dependent expressions
for $=\mean{\dot\phi}$ and ${\cal D}$ we can compute $\tlock/T_0$ as a
function of noise intensity $D$ where $T_0$ is either $1/\gamma$ for
the DMP or $\pi/\Omega$ for the DPP respectively. The results for both
the DMP and the DPP are plotted in Fig.~\ref{fig:tlock}. A pronounced
maximum for intermediate values of noise intensity clearly proves that
noise-induced frequency synchronization is accompanied by
noise-induced phase synchronization.

\section{Oscillatory systems and the Rice frequency}
\label{sec:3}
\subsection{General relations for potential systems}
\label{sec:3:general}
As mentioned in Sec.~\ref{sec:1:phiN} positive-going zero crossings
can be used to count completions of a cycle in oscillatory systems.
In this view the average frequency, i.e., the average phase velocity,
turns out to be the average rate of zero crossings which is captured
by a formula put forward by Rice \cite{Rice44,Rice54}. This elementary
observation yields a novel way to quantify the average frequency of a
phase evolution, henceforth termed the ``Rice frequency'', and to
prove frequency locking in stochastic systems.

To detail our derivation of the Rice frequency in this section, we
start from the following one-dimensional potential system
\begin{equation}
  \ddot x + \gamma\;\dot x + U'(x) = 
  \sqrt{\gamma}\;\xi + F\cos{(\Omega t)}
  \label{e_osc}
\end{equation}
subjected to Gaussian white noise $\xi$ of intensity $D$, i.e.,
\begin{equation}
  \label{eq:noise}
  \langle\xi(t)\rangle = 0,\quad
  \langle\xi(t)\;\xi(s)\rangle=2D\;\delta(t-s)\,,
\end{equation}
and being driven by the external harmonic force $F\cos{(\Omega t)}$.
In \F{f_oscillator} we show a sample path for the harmonic oscillator
\begin{equation} 
  \ddot x + \gamma\;\dot x + \w_0^2\;x = 
  \sqrt{\gamma}\;\xi + F\cos{(\Omega t)}.
  \label{e_harmonicosc} 
\end{equation}
where we used the friction coefficient $\gamma=1$, the natural
frequency $\w_0=1$, and a vanishing amplitude $F=0$ of the external
drive.
As can be read off from \F{f_oscillator}, the velocity $v=\dot x$
basically undergoes a Brownian motion and, therefore, constitutes a
rather jerky continuous, but generally not differentiable signal. In
particular, near a zero crossing of $v$ there are many other zero
crossings. In contrast to that, the coordinate $x$ is a much smoother
signal since it is determined by an integral over a continuous
function
\begin{equation}
  x(t) = x(0) + \int\limits_0^t v(\t) d\t ,
\end{equation}
and, therefore, differentiable. In particular, near a zero crossing of
$x$ there are no other zero crossings. In the following, we will take
advantage of this remarkable smoothness property of $x$ that is an
intrinsic property of the full oscillatory system (\ref{e_osc}) and
disappears when we perform the overdamped limit.

In 1944, Rice \cite{Rice44} deduced a formula for the average number of
zero crossings of a smooth signal like $x$ in the oscillator equation
(\ref{e_osc}). In this rate formula enters the probability density
$P(x,v;t)$ of $x$ and its time derivative, $v=\dot x$, at a given
instant $t$. The Rice rate for passages through zero with positive
slope (velocity) is determined by \cite{Rice54}
\begin{equation} 
  \la f \ra (t) = \int\limits_0^\i v P(x=0,v;t) \, dv .
  \label{e_rice0} 
\end{equation} 
This time-dependent rate is to be understood as an ensemble average.
If the dynamical system is ergodic and mixing the asymptotic
stationary rate $\la f_s \ra$ can likewise be achieved by the
temporal average of a single realization. Let $N([0,t])$ be the number
of positive-going zeros of the signal $x$ in the time interval
$[0,t]$. Using ergodicity, the relation
\begin{equation}
  \label{eq:ergod}
  \la f_s \ra = \int\limits_0^\i v P_s(x=0,v) \, dv  =
  \lim_{t\rightarrow \i} \frac{N\left ([0,t]\right )}{t}
\end{equation}
is fulfilled for the process characterized by the stationary density
$P_s(x,v)$. In the following we always consider stationary quantities.
As explained in Sec.~\ref{sec:1:phiN}, the zero crossings can be
used as marker events to define an instantaneous phase $\phi^L(t)$ by
linear interpolation, cf.~\E{linphase}. The related average phase
velocity is the product of the (stationary) Rice rate and $2\pi$ and,
hence, called the (stationary) Rice frequency
\begin{equation}
  \label{eq:ricefreq}
  \la \w \ra_R = 2\pi\,\la f_s \ra  =
  2\pi\int\limits_0^\i v P_s(x=0,v) \, dv\,.
\end{equation}

For a dynamics described by a potential $U(x)$ in the absence of an
external driving, i.e., (\ref{e_osc}) with $F=0$, the stationary
density can be calculated explicitly yielding
\begin{equation}
  \label{eq:statdens}
  P_s(x,v) = C\,\exp\left[-\left({v^2\over 2}+U(x)\right)/ D\right]
\end{equation}
where $C$ is the normalization constant. From this and the application
of Eq.~(\ref{eq:ricefreq}), it is straightforward to derive the exact
result
\begin{equation}
  \label{eq:potsys}
  \la \w \ra_R = \frac{\sqrt{2\pi D}\,\exp\left[-{U(0)\over D}\right]}
  {\int\limits_{-\infty}^{\infty}\exp\left[-{U(x)\over D}\right]\,dx}\,.
\end{equation}
Without loss of generality we can set $U(0)=0$. In the limit $D\to 0$,
we can perform a saddlepoint approximation around the deepest minima
$x_m$ (e.g.~for symmetric potentials). In this way we find the
following expression valid for $D\ll \Delta U=U(0)-U(x_i)$ , i.e., the
small noise approximation,
\begin{equation}
 \label{eq:smallnoise}
 \la \w \ra_R = \left[
   \sum_i\,{\exp\left[-{U(x_i)\over D}\right]\over\sqrt{U^{''}(x_i)}}
   \right]^{-1}\,.
\end{equation}
In the limit $D\to\infty$, we have to consider the asymptotic behaviour
of the potential, $\lim_{x\to\pm\infty}U(x)$, to estimate the integral
in Eq.~(\ref{eq:potsys}). For potentials that can be expanded in a
Taylor series about zero and that, therefore, result in a power series
of order $2m$, i.e., $U(|x|\to\infty)\sim x^{2m}$, we can rescale the
integration variable by $x=D^{1/2m}\tilde x$.  For sufficiently large
$D$, the integral is dominated by the power $2m$ term. In this way we
find the large noise scaling
\begin{equation}
 \label{eq:largenoise}
 \la \w \ra_R \;\stackrel{\tiny(D\to\infty)}{\sim}\; D^{\alpha},
 \quad\mbox{with}\quad
 \alpha={m-1\over 2m}\,.
\end{equation}
Applying Eqs.~(\ref{eq:potsys}) and (\ref{eq:smallnoise}) to the
harmonic oscillator (\ref{e_harmonicosc}) we immediately find that
$\la \w \ra_R = \w_0$, independent of $\gamma$ and for all values of
$D>0$.  This is also in agreement with Eq.~(\ref{eq:largenoise}). It
follows because $m=1$ implies that, for large noise, the Rice
frequency $\la \w \ra_R$ does not depend on $D$ at all. Note, however,
that in the deterministic limit, i.e., for $D=0$, we have the standard
result
\begin{equation}
  \label{eq:disco}
  \la \w \ra_R  \;\stackrel{\tiny(D=0)}{=}\; \left\{ 
    {\sqrt{\w_0^2-\g^2/4}\quad\mbox{for } \g<2\w_0
      \atop
      \qquad 0 \qquad\qquad\mbox{for } \g\ge 2\w_0}
    \right.\,,
\end{equation}
which explicitly does depend on the friction strength $\g>0$.
Therefore, the limit $D=0$ is discontinuous except in the undamped
situation $\g=0$.

The similarity of Eqs.~(\ref{eq:potsys}) and (\ref{eq:smallnoise})
with rates from transition state theory \cite{HangTalkBork90} will be
addressed below when we discuss the bistable potential.

\subsection{ The role of coloured noise}
\label{sec:3:colnoi}
It is well known that the Rice frequency cannot be defined for
stochastic variables that integrate increments of the Wiener process
(white noise).
From Eq.~(\ref{e_osc}) this holds true for the velocity $\dot v=\ddot
x$. This is so, because the stochastic trajectories of degrees of
freedom being subjected to Gaussian white noise forces are continuous
but are of {\em unbounded} variation and nowhere differentiable
\cite{book,HanggiThom82}. This fact implies that such stochastic
realizations cross a given threshold within a fixed time interval
infinitely often if only the numerical resolution is increased {\em ad
  infinitum}. This drawback, which is rooted in the mathematical
peculiarities of idealized Gaussian white noise, can be overcome if we
consider instead a noise source possessing a finite correlation time,
i.e., coloured noise, see ref.~\cite{HJ95}. To this end, we consider
here an oscillatory noisy harmonic dynamics driven by Gaussian
exponentially correlated noise $z(t)$, i.e.,
\begin{eqnarray}
  \label{eq:col1a}
  \dot x &=& v\\
  \label{eq:col1b}
  \dot v &=& -\g \;v -\w_0^2 \;x + \sqrt{\g}\;z(t)\\
  \label{eq:col1c}
  \dot z &=& -\frac{z}{\tau} + \frac{1}{\tau}\;\xi\,,
\end{eqnarray}
with $z(t)$ obeying $\langle z(t)\rangle=0$ and
\begin{equation}
  \label{eq:col2}
  \langle z(t)\;z(s)\rangle = \frac{D}{\tau} \exp\left(
    -\frac{|t-s|}{\tau}
  \right)\,.
\end{equation}
Following the same reasoning as before we find for the Rice
frequency of $x(t)$ as before
\begin{eqnarray}
  \la\w\ra_x 
  &=&
  \int\limits_0^{\infty}dv\int\limits_{-\infty}^{\infty} dz
  \;v\,P_s(0,v,z)\\
  &=&
  \frac{\w_0}{\sqrt{1+\g\tau}}\,.
  \label{eq:wx}
\end{eqnarray}

Likewise, upon noting that within a time interval $\Delta t, -\Delta
t(-\g\dot x -\w_0^2 x +\sqrt{\g} z)<v<0$, or $-\Delta t (-\w_0^2 x
+\sqrt{\g}z) + {\cal O}(\Delta t)^2<v<0$, respectively, the Rice
frequency of the zero crossings with positive slope of the process
$v(t)$ is given by
\begin{equation}
\label{eq:wv}
  \la\w\ra_v 
  =
  \int\limits_{-\infty}^{\infty}dx\int\limits_x^{\infty} dz\;
  (\sqrt{\g}\;z-\w_0^2\;x)\,P_s(x,0,z)\,,  
\end{equation}
which is evaluated to read
\begin{equation}
  \la\w\ra_v 
  =
  \sqrt{\w_0^2+\frac{\g}{\tau}}\,.
\end{equation}
The result in (\ref{eq:wx}) shows that for small noise colour $\tau$
the Rice frequency for $\langle\w\rangle_x$ assumes a correction
$\langle\w\rangle_x\sim\w_0\left(1-\frac{\tau}{2\g}\right)$, as
$\tau\rightarrow 0^+$. In clear contrast, the finite Rice frequency
for the velocity process $v(t)$ (\ref{eq:col1b}) diverges in the limit
of vanishing noise colour proportional to $\tau^{-1/2}$.

\subsection{Relation between Rice and Hilbert frequency}
\label{sec:3:ricehilbert}
To exemplify the relation between the Rice frequency $\la \w \ra_R$
and the Hilbert frequency $\la \w \ra_H$, again we consider the damped
harmonic oscillator \E{e_harmonicosc} agitated by noise alone. In
\F{f_hilbertoscillator} we show a numerically evaluated sample path
and the corresponding Hilbert phase (normalized to $2\pi$ and modulo
$1$) using the parameters $\gamma=1, D=1, \w_0=1, F=0$.
\begin{figure}
  \centerline{\epsfig{figure=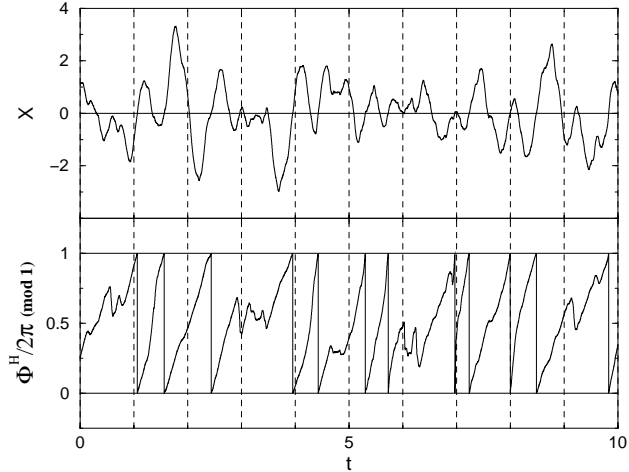,width=\figwidth}}
  \caption{Signal $x(t)$ (top panel) and corresponding Hilbert phase
    $\phi^H(t)/2\pi$ modulo 1 (bottom panel) for the undriven harmonic
    oscillator \E{e_harmonicosc} with friction strength $\gamma=1$,
    noise intensity $D=1$, natural frequency $\w_0=1$, and driving
    amplitude $F=0$. Note that although there are successive zero
    crossings of $x$ with positive slope near $t\approx 3$ and
    $t\approx 9$ the Hilbert phase does not increase by $2\pi$.}
  \label{f_hilbertoscillator}
\end{figure}
An important point to observe here is that around $t\approx 3$ and
$t\approx 9$ the Hilbert phase $\phi^H$ does not increase by $2\pi$
after two successive passages through zero with positive slope. This
shall illustrate the difference between the Hilbert phase and the
natural phase. In Subsec.~\ref{sec:1:phiH} this observation was
already mentioned as a consequence of the nonlocal character of the
Hilbert transform. In particular, short and very small amplitude
crossings to positive $x$ are not properly taken into account by the
Hilbert phase since they only result in a small reduction of $\phi^H$.
This leads us to conjecture that quite generally
\begin{equation}
  \la  \w \ra_R \ge  \la  \w \ra_H
  \label{e_ungleich}
\end{equation}
holds. In fact, for the case of the harmonic oscillator that generates
a stationary Gaussian process one even can prove this conjecture by
deriving explicit expressions for $\la \w \ra_R$ and $\la \w \ra_H$.
As usual, let $S(\omega)$ denote the spectrum of the stationary
Gaussian process $x$. Then the Rice frequency can be recast in the
form of \cite{Rice54}
\begin{equation}
  \la  \w \ra_R  =
  \left [
    \frac{\int\limits_0^\i \w^2 \, S(\w) \; d\w}
    {\int\limits_0^\i  S(\w) \; d\w}
  \right ]^{1/2}\,.
  \label{e_srice}
\end{equation}
A similar expression (additionally involving an Arrhenius-like
exponential) exists when considering not zero crossings, as in
Eq.~(\ref{e_srice}), but crossings of an arbitrary threshold. In
\cite{Callenbach} it was shown that the Hilbert frequency of the same
process $x$ is given by a similar expression, namely
\begin{equation}
  \la  \w \ra_H =
  \left [
    \frac{\int\limits_0^\i \w \, S(\w) \; d\w}
    {\int\limits_0^\i  S(\w) \; d\w}
  \right ]\,.
  \label{e_shil}
\end{equation}
Interpreting the quantity $S(\w)/\int_0^\infty S(\hat\w) d\hat\w$ as a
probability density $P(\w)$, $\w \in (0,\infty)$, we can use the
property that the related variance is positive, i.e.,
\begin{equation}
  \label{eq:var}
  \int\limits_0^{\infty} \w^2 P(\w)\; d\w\ge
  \left [ \int\limits_0^{\infty} \w P(\w)\; d\w \right ] ^2 \,.
\end{equation}
Taking the square-root on both sides of the last inequality
immediately proves \E{e_ungleich}.

Using the spectrum of the undriven noisy harmonic oscillator
\begin{equation}
  S(\omega)=\frac{4\gamma D}{(\w_0^2-\w^2)^2+\gamma^2\w^2}\,
\end{equation}
and employing Eqs.~(\ref{e_srice}) and (\ref{e_shil}), it is easy to
see that both $\la\w\ra_R$ and $\la\w\ra_H$ do not vary with $D$. We
have already shown above that $\la\w\ra_R=\w_0$. In contrast to this,
$\la\w\ra_H$ is a monotonically decreasing function of $\gamma$ that
approaches $\w_0$ from below in the limit $\gamma\to 0^+$.

\subsection{Periodically driven noisy harmonic oscillator}
\label{sec:3:harmosciplusdrive}
The probability density of the periodically driven noisy harmonic
oscillator can be determined analytically by taking advantage of the
linearity of the problem. Introducing the mean values of the coordinate
and the velocity, $ \la x(t)\ra$ and $\la v(t)\ra $, the variables
\begin{equation} 
  \tilde x=x-\la x\ra,  \qquad \qquad \tilde v=v-\la v\ra
\end{equation} 
obey the differential equation of the undriven noisy harmonic
oscillator. In the asymptotic limit $t\rightarrow \infty$ the mean
values converge to the well known deterministic solution
\begin{eqnarray}
  \la x(t) \ra &=& 
  \frac{F}{\sqrt{ (\w_0^2-\Omega^2)^2+\gamma^2\Omega^2}}
  \;\cos( \Omega t-\delta)\\
  \la v(t) \ra &=& 
  -\Omega \frac{F}{\sqrt{ (\w_0^2-\Omega^2)^2+\gamma^2\Omega^2}}
  \;\sin( \Omega t-\delta) \\
  \delta &=& \arctan\left[\frac{\gamma \Omega}{\w_0^2-\Omega^2}\right] 
\end{eqnarray}
with the common phase lag $\delta$. Therefore, after deterministic
transients have settled the {\em cyclo-stationary} probability density
of the driven oscillator reads
\begin{equation}
  P_{cs}(x,v;t) = P_s(x-\la x(t)\ra,v-\la v(t)\ra) 
  \label{e_pharm}
\end{equation}
with the Gaussian density 
\begin{equation}
  P_s(x,v) = \frac{\w_0}{2 \pi D} \exp {\left [ 
      - \left (\frac{v^2}{2}+\frac{\w_0^2 x^2}{2} \right )/D  \right ] } .
\end{equation} 
Using Eq.~(\ref{eq:ricefreq}) the cyclo-stationary probability density
(\ref{e_pharm}) yields an oscillating expression for the Rice
frequency $\la\w\ra_{\!R}\,(t)$.  The time dependence of this
stochastic average can be removed by an initial phase average, i.e., a
subsequent average over one external driving period $2\pi / \Omega$,
\begin{eqnarray}
  \la \w\ra_R &=& 
  \int\limits_0^{2\pi/\Omega}\,\la\w\ra_{\!R}\;(t)\;\frac{\Omega dt}{2\pi}\\
  &=& 
  \int\limits_0^{2\pi/\Omega}\,\int\limits_0^\infty v\, 
  P_{cs}(0,v;t)\; dv\;\Omega dt\,. 
  \label{e_drivenosc} 
\end{eqnarray}
The resulting analytical and numerically achieved values of the Rice
frequency as a function of the noise intensity $D$ are shown in
\F{f_oscillatordriven} for fixed $\w_0=1,F=1,\Omega=3$ and various
values of $\gamma$.
\begin{figure}
  \centerline{\epsfig{figure=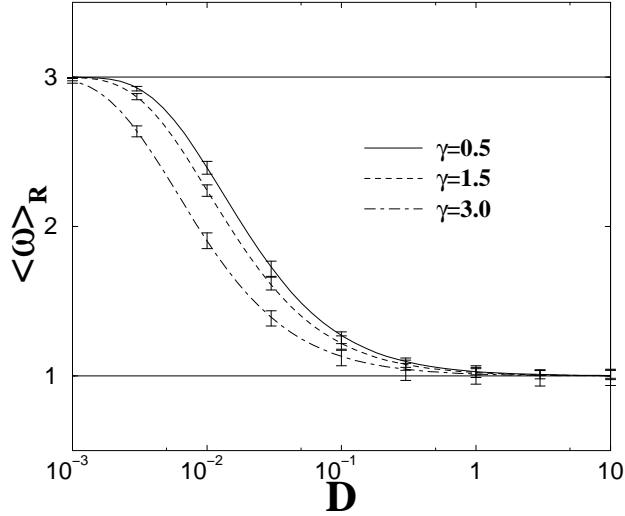,angle=-90,width=\figwidth}}
  \caption{Rice frequencies for the driven harmonic oscillator
    \E{e_harmonicosc} with natural frequency $\w_0=1$, driving
    amplitude $F=1$, and driving frequency $\o=3$ for different values
    of the friction strength $\gamma$. The numerically achieved values
    (symbols with error bars) match the analytical curves determined
    using \E{e_drivenosc}.}
  \label{f_oscillatordriven}
\end{figure}
For small noise intensities $D$ the Rice frequency $\la\w\ra_R$ is
identical to the external driving frequency $\Omega$, whereas for
large noise intensities the external drive becomes inessential and the
Rice frequency approaches $\la\w\ra_R=\w_0$.

Further insight into the analytic expression (\ref{e_drivenosc}) is
gained from performing the following scale transformations
\begin{equation}
  \label{eq:scaletx}
  \tilde t = \Omega \,t -\delta
  \qquad\mbox{and}\qquad
  \tilde x = \frac{x}{\sqrt{2D}/\Omega}
\end{equation}
from which we immediately find the rescaled velocity
\begin{equation}
  \label{eq:scalev}
  \tilde v = \frac{d\tilde x}{d\tilde t} =
  \frac{\sqrt{2D}/\Omega}{1/\Omega}\frac{dx}{dt} = \sqrt{2D}\;v\,.
\end{equation}
Inserting these dimensionless quantities into Eq.~(\ref{e_drivenosc}) yields
\begin{eqnarray}
  \la\w\ra_R &=& \w_0\;I(\tilde A,\tilde\w_0)\\[.3cm]
  I(\tilde A,\tilde\w_0)&=&{1\over\pi}
  \int\limits_{-\delta}^{2\pi-\delta}\;\int\limits_0^{\infty}
  \tilde v\,\exp\Big[\!\!
    -(\tilde v + \tilde A \sin\tilde t)^2
  \nonumber\\[-.3cm]
  &&\qquad\qquad\qquad\qquad
    -(\tilde\w_0\,\tilde A \cos\tilde t)^2
    \Big]\,d\tilde v\;d\tilde t
    \label{eq:i}
\end{eqnarray}
where we have defined further dimensionless quantities
\begin{eqnarray}
  \tilde A &=& \frac{\Omega}{\sqrt{2D}}
  \frac{F}{\sqrt{
      (\w_0^2- \Omega^2)^2 + (\gamma \Omega)^2
      }}\\[.3cm]
  \tilde\w_0 &=& \frac{\w_0}{\Omega} .
\end{eqnarray}
Due to the $2\pi$ periodicity of the trigonometric functions, the
integral (\ref{eq:i}) does not change when shifting the interval for
the  integration with respect to $\tilde t$ back to $[0,2\pi]$.
Hence, $I$ is only a
function of $\tilde A$ and $\tilde\w_0$. An expansion for small
$\tilde A$ yields
\begin{equation}
  \label{eq:A20}
  \la\w\ra_R=
  \w_0\left[1+\frac{1-\tilde\w_0^2}{2}\tilde A^2 + 
    {\cal O}(\tilde A^4)\right]
\end{equation}
which implies for large $D/F^2$
\begin{equation}
  \label{eq:D2inf}
  \la\w\ra_R-\w_0\;
  \sim\;\frac{F^2}{D}\,.
\end{equation}
The opposite extreme, $\tilde A\to\infty$ or $D/F^2\to 0$, can be
extracted from a saddlepoint approximation around $\tilde v=\tilde A$
and $\tilde t = 3\pi/2$. Following this procedure, the integral
(\ref{eq:i}) gives the constant $1/\tilde\w_0$. This directly implies
$\la\w\ra_R=\Omega$.

The crossover between these two extremes occurs when the first
correction term in (\ref{eq:A20}) is no longer negligible, i.e., for
\begin{equation}
  \frac{|1-\tilde\w_0^2|}{2}\tilde A^2\approx 1.
\end{equation}
When solved for the crossover noise intensity $D_{\rm co}/F^2$, this
yields
\begin{equation}
  \frac{D_{\rm co}}{F^2} \approx \frac{|\Omega^2-\w_0^2|}
  {4\left[(\w_0^2- \Omega^2)^2 + (\gamma \Omega)^2\right]}\,,
\end{equation}
which, for the parameters used in \F{f_oscillatordriven}, correctly gives
values between $10^{-2}$ and $10^{-1}$.

In \F{f_oscillatordriven} the parameters $F,\o$, and $\w_0$ and, hence,
$\tilde\w_0$ are identical for all curves. Solving $\tilde
A(\g_1,D_1)=\tilde A(\g_2,D_2)$ with respect to $D_2$ shows that the
curves become shifted horizontally as in the log-linear plot in
\F{f_oscillatordriven}. Another way to explain this shift is by
noting that $dD_{\rm co}/d\gamma <0$.

\section{Bistable Kramers oscillator: noise-induced phase coherence and SR}
\label{sec:4}
\subsection{Rice frequency and transition state theory}
\label{sec:4:}
The bistable Kramers oscillator, i.e., Eq.~(\ref{e_osc}) with the
double well potential
\begin{equation} 
  U(x)= \frac{x^4}{4}-\frac{x^2}{2}\,,
  \label{e_potbiosc}
\end{equation} 
is often used as a paradigm for nonlinear systems. With reference to
\E{e_osc} the corresponding Langevin equation is given by
\begin{equation}
  \ddot x + \gamma \dot x +  x^3-x = 
  \sqrt{\gamma} \xi\,+\, F\cos{(\Omega t)}\,
  \label{e_biosc}
\end{equation}
which, in the absence of the external signal, $F=0$, generates the
stationary probability distribution
\begin{equation}
  P_s(x,v) = C \exp \left \{ -\left ( \frac{v^2}{2}
      + \frac{x^4}{4}-\frac{x^2}{2} \right ) /D  \right \}
  \label{e_probdensbi}
\end{equation}
with the normalization constant $C$. Using this stationary probability
density and Eq.~(\ref{eq:ricefreq}) we can determine the Rice
frequency analytically. In \F{f_bioscstatic} we depict this analytic
result together with numerical simulation data including error bars.
\begin{figure}
  \centerline{\epsfig{figure=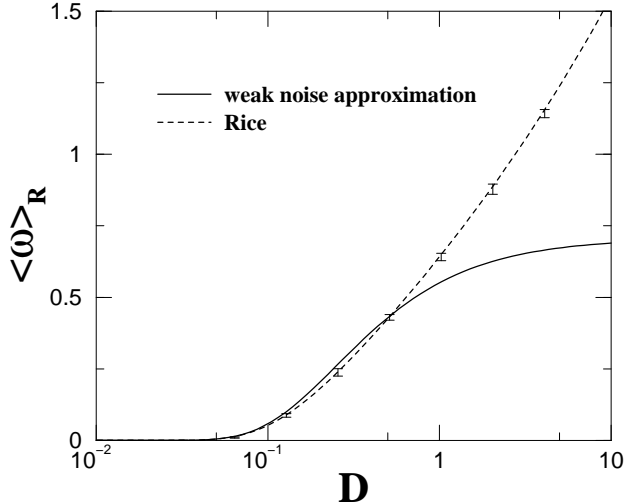,angle=-90,width=\figwidth}}
  \caption{Rice frequencies for the undriven bistable oscillator
    \E{e_biosc} with friction strength $\gamma=1$. Numerical values
    with error bars match the analytically determined values (dotted
    line) using \E{eq:ricefreq} with \E{e_probdensbi}. As expected,
    for large values of $D$ the Rice frequency scales like $D^{1/4}$.
    The solid line presents the leading weak noise approximation in
    Eq.~(\ref{eq:TSTRAPROX}).}
  \label{f_bioscstatic}
\end{figure}
The simulation points perfectly match the analytically determined
curve. As expected for the asymptotically dominant quartic term, i.e.,
$m=2$ (cf.~Sec.\ref{sec:3:general}, especially \E{eq:largenoise}), the
Rice frequency scales as $\la \w\ra_R \sim D^{1/4}$ for large values
of $D$ .

Comparing the Rice frequency formula, Eq.~(\ref{eq:ricefreq}), with
the forward jumping rate $k^+_{\rm TST}$ from the transition state
theory \cite{HangTalkBork90},
\begin{equation} 
  k^+_{\rm TST} = 
  \hat Z_0^{-1}\int dx\,dv\;\theta(v)\,\delta (x)\;v\;
  \exp\left[-H(x,v)/D\right]
  \label{eq:tstint}
\end{equation}
where
\begin{equation} 
  \hat Z_0=\int\limits_{x<0} dx\,dv\;\exp\left[-H(x,v)/D\right]\,,
  \label{eq:semipf}
\end{equation}
and $H(x,v)= (1/2)v^2+(1/4)x^4-(1/2)x^2$ represents the corresponding
Hamiltonian, one can see that the difference between both solely rests
upon normalizing prefactors. Whereas the rate $k^+_{\rm TST}$ is
determined by the division of the integral \E{eq:tstint} by the
``semipartition'' function $\hat Z_0$, the rate $\la \w\ra_R/2\pi$ is
established by dividing the same integral \E{eq:tstint} by the
complete partition function $Z_0$
\begin{equation} 
  Z_0=\int\limits dx \,dv\,  \exp\left[-H(x,v)/D\right]\,.
  \label{eq:fullpf}
\end{equation}
Particularly for symmetric (unbiased) potentials, i.e., $V(-x)=V(x)$,
this amounts to the relation $Z_0=2 \hat Z_0$, hence,
\begin{equation}
  \la\w\ra_R = \pi\, k^+_{\rm TST}.
  \label{eq:TSTR}
\end{equation}
At weak noise, $E_b/D >> 1$, this relation simplifies to
\begin{equation}
  \la\omega\ra_R \approx {\omega_{0} \over 2} \exp [- E_b/D] \;,
  \label{eq:TSTRAPROX}
\end{equation}
wherein $E_b$ denotes the barrier height and $\w_0$ the angular
frequency inside the well ($\w_0=\sqrt{2}$). Indeed, in the
small-to-moderate regime of weak noise this estimate nicely predicts
the exact Rice frequency (cf.~Fig.~\ref{f_bioscstatic}).

\subsection{Periodically driven bistable Kramers oscillator}
\label{sec:4:perdrivkram}
The periodically driven bistable Kramers oscillator was the first
model considered to explain the phenomenon of SR \cite{SR_hist} and it
still serves as one of the major paradigms of SR
\cite{SR_rev1,SR_rev2}. In its overdamped form it was used to support
experimental data (from the Schmitt trigger) displaying the effect of
stochastic frequency locking \cite{nei_lsg98,shulgin} observed for
sufficiently large, albeit subthreshold signal amplitudes, i.e., for
$F_{\rm min}<F<2/\sqrt{27}$. From a numerical simulation of the
overdamped Kramers oscillator and computing the Hilbert phase it was
also found that noise-induced frequency locking for large signal
amplitudes was accompanied by noise-induced phase coherence, the
latter implies a pronounced minimum of the effective phase diffusion
coefficient
\begin{equation} 
  {\tilde D}_{\rm eff} = \;
  \frac{1}{2}\partial_t\left[\langle(\phi(t))^2\rangle
    -\langle\phi(t)\rangle^2\right]
  \label{eq:deff1}
\end{equation}
occurring for optimal noise intensity. Based on a discrete model
\cite{NeiLsgMoShulCol99}, analytic expressions for the frequency and
phase diffusion coefficient were derived that correctly reflect the
conditions for noise-induced phase synchronization
\cite{FreuNeiLsg2000} for both periodic and aperiodic input signals.

To link the mentioned results to the Rice frequency introduced above
we next investigate the behaviour of the Kramers oscillator with
non-vanishing inertia. We show numerical
simulations for \E{e_biosc} with the parameters $\o=0.01,\gamma=0.5$
and diverse values of $F$ in \F{f_nonlinearF}.
\begin{figure}
  \centerline{\epsfig{figure=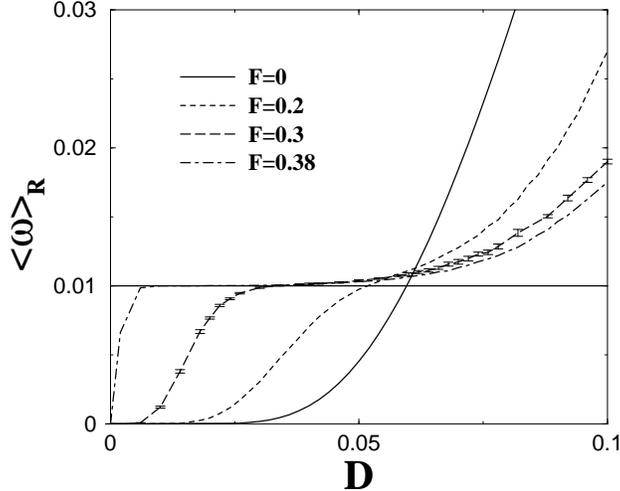,angle=-90,width=\figwidth}}
  \caption{Numerically determined Rice frequencies of the periodically
    driven bistable Kramers oscillator \E{e_biosc} computed with the
    friction coefficient $\gamma=0.5$ and the angular driving
    frequency $\o=0.01$ and plotted as a function of the noise
    intensity $D$. Different curves correspond to various amplitudes
    of the harmonic drive $F$. For larger values of $F$ wider regions
    appear where the Rice frequency is locked to the external driving
    frequency $\Omega$.}
  \label{f_nonlinearF}
\end{figure}
For larger values of $F$, a region around $D\approx 0.05$ appears where
the Rice frequency is locked to the external driving frequency
$\Omega$. Since for larger values of the external driving $F$ smaller
values of the noise parameter $D$ are needed to obtain the same rate for
switching events, the entry into the locking region shifts to smaller
values of $D$ for increasing $F$.

In \F{f_nonlinearrice1} we present numerical simulations for fixed
$F=0.384,\o=0.01$ and different values of the damping coefficient
$\gamma$. Note that the value of $F$ is slightly smaller than the
critical value $F_{\rm c} = 2/\sqrt{27}\approx 0.3849...$. For smaller
values of $\gamma$ wider coupling regions appear since it is easier
for the particle to follow the external driving for smaller damping.
\begin{figure}
  \centerline{\epsfig{figure=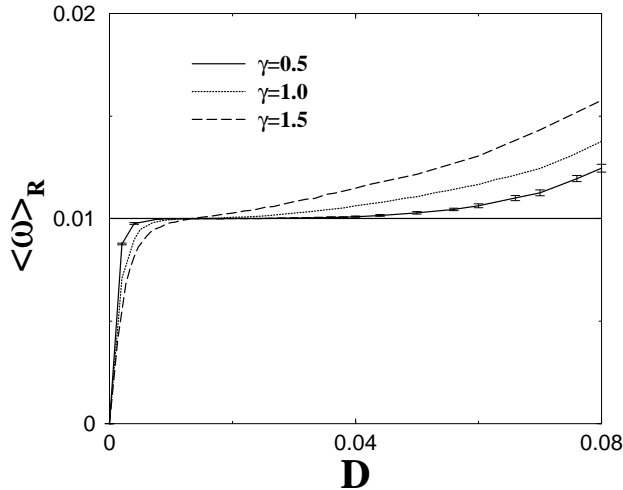,angle=-90,width=\figwidth}}
  \caption{Numerically determined Rice frequency as a function of
    the noise intensity $D$ for the periodically driven Kramers
    oscillator \E{e_biosc} with the angular driving frequency
    $\o=0.01$ and driving amplitude $F=0.384$ for different values of
    the friction coefficient $\gamma$. For smaller values of $\gamma$
    wider regions of frequency locking appear.}
  \label{f_nonlinearrice1}
\end{figure}

To check whether frequency synchronization is accompanied by effective
phase synchronization we have also computed the averaged effective
phase diffusion coefficient, this time defined by the following
asymptotic expression
\begin{equation}
  \label{eq:deff2}
  D_{\rm eff} = \lim_{t\rightarrow \infty} \frac{1}{2t}
    \left\langle
    \left [ \phi(t) -\langle\phi(t)\rangle \right ]^2
  \right\rangle\,.
\end{equation}
It should be clear that the instantaneous ``Rice'' phase $\phi(t)$ was
determined via zero crossings. The connection with the instantaneous
diffusion coefficient defined in (\ref{eq:deff1}) is established by
applying the limit $t\to\infty$
\begin{equation}
  \label{eq:deff3}
  D_{\rm eff} = \lim\limits_{t\to\infty}\;
  \frac{1}{t}\;
  \int\limits_0^t  {\tilde D}_{\rm eff}(\tilde t)\;
  d\tilde t\,.
\end{equation}
In \F{f_nonlineardiff1} we show numerical simulations of the effective
phase diffusion coefficient $D_{\rm eff}$ as function of noise
intensity $D$.
\begin{figure}
  \centerline{\epsfig{figure=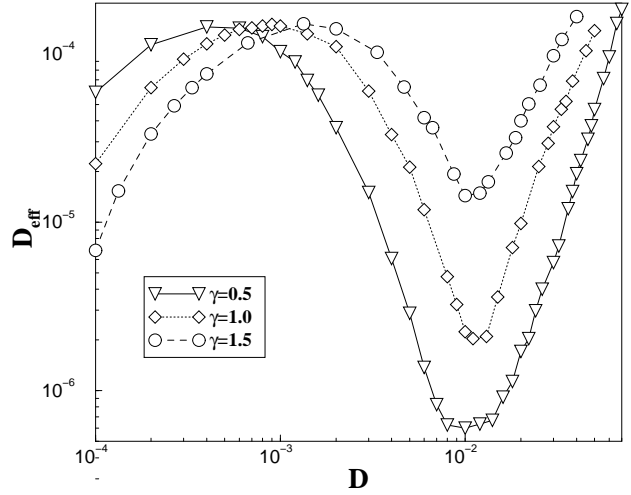,width=\figwidth}}
  \caption{Effective phase diffusion coefficient {\em vs}.~noise
    intensity for the periodically driven bistable Kramers oscillator
    \E{e_biosc} with angular driving frequency $\o=0.01$, driving
    amplitude $F=0.384$, which is close-to-threshold driving, and for
    different values of $\gamma$. For smaller values of the friction
    coefficient $\gamma$ phase diffusion is diminished.}
  \label{f_nonlineardiff1}
\end{figure}
The phase diffusion coefficient displays a local minimum that gets
more pronounced if the damping coefficient $\gamma$ is decreased.
Indeed, phase synchronization reveals itself through this local
minimum of the average phase diffusion coefficient $D_{\rm eff}$ in
the very region of the noise intensity $D$ where we also observe
frequency synchronization, cf.~\F{f_nonlinearF}.
\begin{figure}
  \centerline{\epsfig{figure=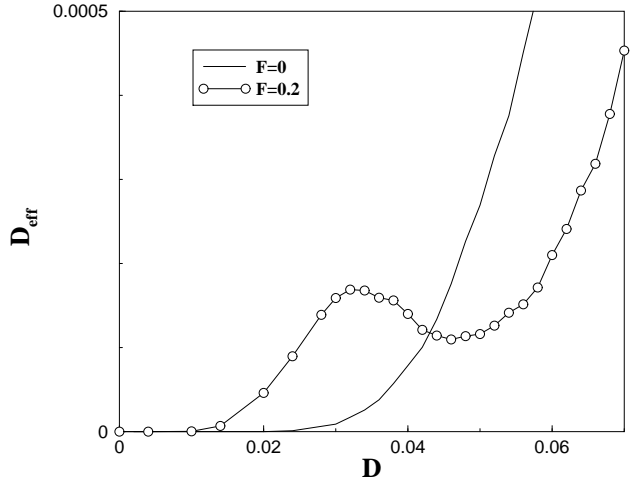,width=\figwidth}}
  \caption{Effective phase diffusion coefficient {\em vs}.~noise
    intensity for the periodically driven bistable Kramers oscillator
    \E{e_biosc} with friction coefficient $\gamma=0.5$, angular
    driving frequency $\o=0.01$, plotted for the undriven case $F=0$
    and for driving with an amplitude $F=0.2$.}
  \label{f_nonlineardiff1a}
\end{figure}
The qualitative behaviour of the diffusion coefficient agrees also with
a recently found result related to diffusion of Brownian particles in
biased periodic potentials \cite{fnl}. A necessary condition for the
occurrence of a minimum was an anharmonic potential in which the
motion takes place. In this biased anharmonic potential the motion
over one period consists of a sequence of two events. Every escape
over a barrier (Arrhenius-like activation) is followed by a time scale
induced by the bias and describing the relaxation to the next minimum.
The second step is weakly dependent on the noise intensity and the
relaxation time may be even larger then the escape time as a result of
the anharmonicity. For such potentials the diffusion coefficient
exhibits a minimum for optimal noise, similar to the one presented in
Figs.~\ref{f_nonlineardiff1} and \ref{f_nonlineardiff1a}.

The average duration of locking episodes $\la T_{\rm lock}\ra$ can be
computed by equating the second moment of the phase difference
(between the driving signal and the oscillator) to $\pi^2$
\cite{FreuNeiLsg2001}. A rough estimate, valid for the regions where
frequency synchronization occurs, i.e., where the dynamics of the
phase difference is dominated by diffusion, thus reads $\la T_{\rm
  lock}\ra =\pi^2/D_{\rm eff}$ or, when expressed by the number of
driving periods \cite{pfish}
\begin{equation}
  \label{eq:nlock}
  \la n_{\rm lock}\ra = {\o\;\pi\over 2\; D_{\rm eff}}\,.  
\end{equation}
In this way we estimate from Figs.~\ref{f_nonlineardiff1} and
\ref{f_nonlineardiff1a} $\la n_{\rm lock}\ra\sim 150\ldots 15000$ for
$\Omega=0.01$ and relevant $D_{\rm eff}$ varying between
$10^{-4}\ldots10^{-6}$.

\subsection{SR without noise-enhanced phase coherence}
\label{sec:4:SRnosync}
\begin{figure}
  \centerline{\epsfig{figure=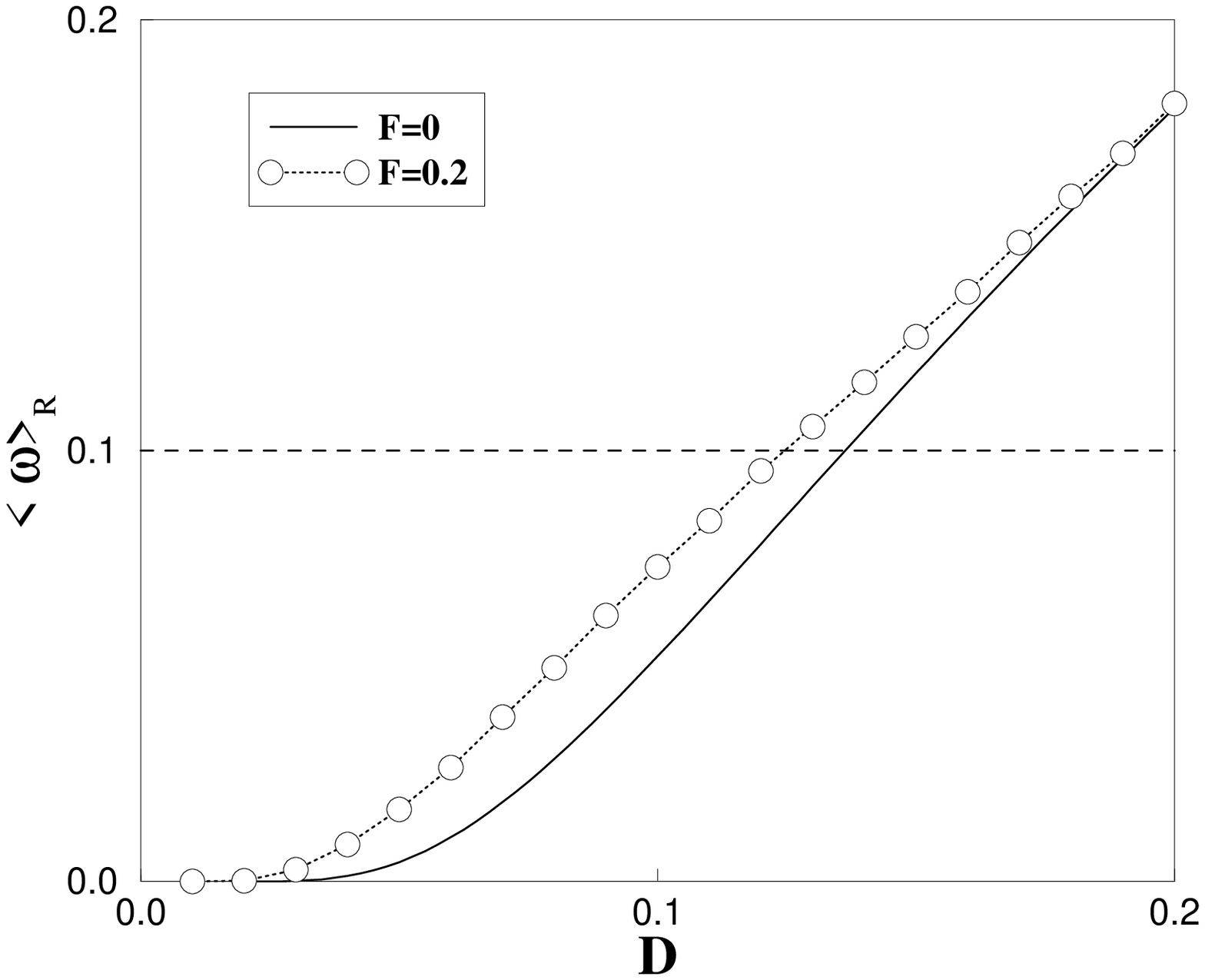,width=\figwidth}}  
  \centerline{\epsfig{figure=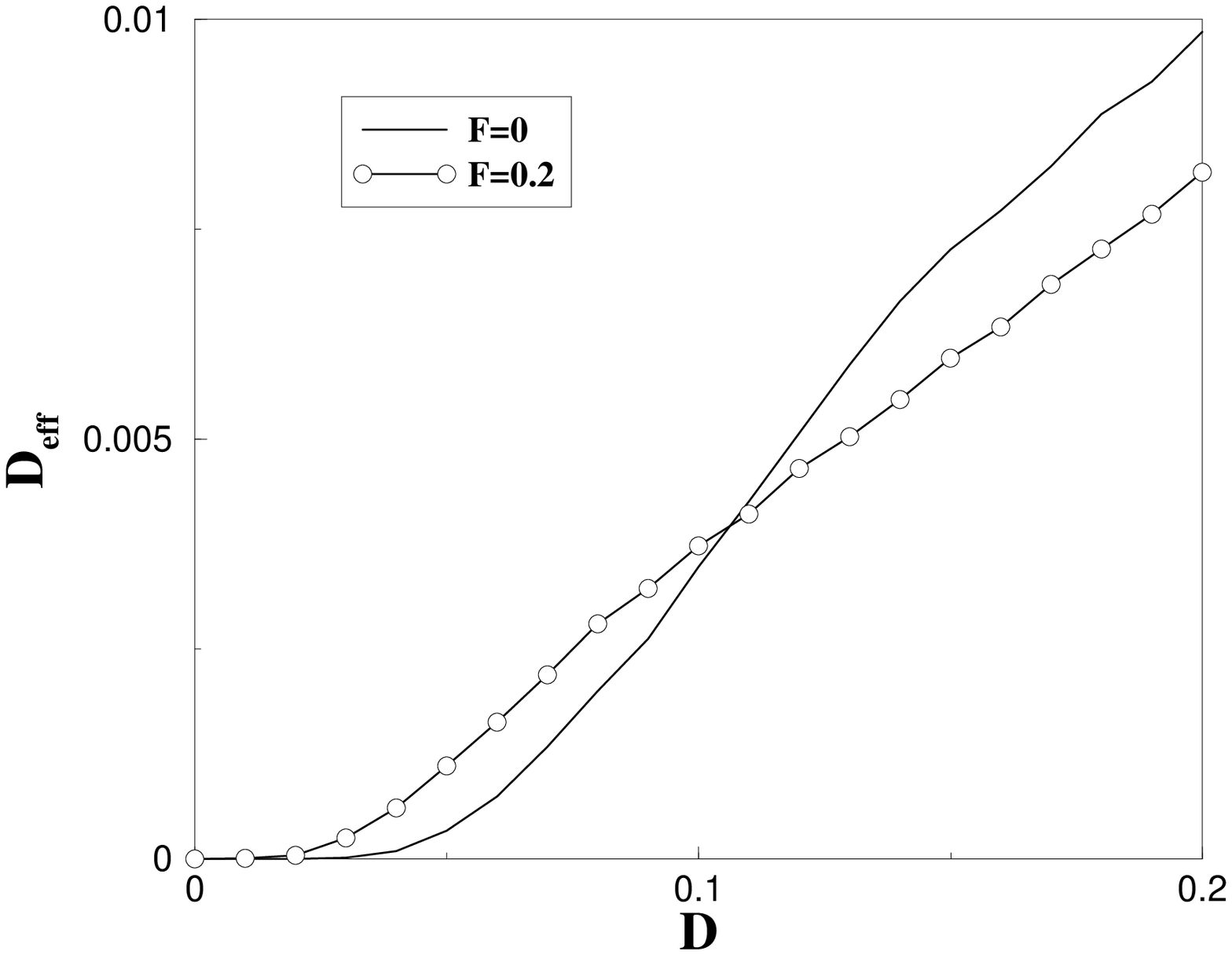,width=\figwidth}}  
  \caption{For friction coefficient $\gamma=1$ and external driving
    frequency $\Omega=0.1$ the bistable Kramers oscillator does
    neither exhibit frequency synchronization nor noise-enhanced phase
    coherence but still stochastic resonance occurs for noise
    intensities in the range of values ($D\approx 0.15$) where
    $\la\w\ra_R\approx \Omega$.}
  \label{fig:SRnosync}
\end{figure}

In the previous examples we have shown how frequency synchronization,
revealing itself through a plateau of the output frequency matching
the harmonic input frequency, and reduced phase diffusivity together
mark the occurrence of noise-enhanced phase coherence. Optimal noise
intensities were found in the range where one also observes SR (in the
overdamped system). In order to underline that under certain
conditions SR exists but may not be accompanied by effective phase
synchronization we present simulation results for the Rice frequency
and the diffusion coefficient in Fig.~\ref{fig:SRnosync} obtained for
the bistable Kramers oscillator with a friction coefficient $\gamma=1$
and external frequency $\Omega=0.1$.
For noise intensities $D\approx 0.15$ the output frequency matches
$\Omega$ and nearby the overdamped Kramers oscillator exhibits the
phenomenon of SR, i.e., one finds a maximum of the spectral power
amplification \cite{SR_rev1,SR_rev2}. In contrast, we neither can find
a minimum in the diffusion coefficient nor a plateau around $\Omega$
meaning that no phase coherence and not even frequency synchronization
can be observed. The reason is that the external signal switches much
too fast for the bistable system to follow; note that in the two-state
description with Arrhenius rates the prefactor $\alpha_0$
(cf.~Eq.~\ref{a1a2}) restricts the switching frequency from above.
Noise-induced phase coherence requires a device with a faster internal
dynamics, i.e., $\Omega\ll \alpha_0$.
\vspace*{3mm}
\section{Conclusions}
\label{sec:concl}
We underline that the noise-induced phase synchronization is a much
more stringent effect than stochastic resonance. This statement
becomes most obvious when recalling that the spectral power
amplification attains a maximum at an optimal noise intensity for
arbitrarily small signal amplitudes and any frequency of the external
signal. In contrast, noise-induced phase synchronization and even
frequency locking are nonlinear effects and as such require
amplitude and frequency to obey certain bounds (see the ``Arnold
tongues'' in Sec.~\ref{sec:2}). We expect that the functioning of
important natural devices, e.g.~communication and information
processing in neural systems or sub-threshold signal detection in
biological receptors, rely on phase synchronization rather than
stochastic resonance.
\acknowledgements The authors acknowledge the support of this work by
the Deutsche Forschungsgemeinschaft, Sfb 555 ``Komplexe Nichtlineare
Prozesse'', project A 4 (L.S.-G.~and J.F.), SFB 486 ``Manipulation of
matter on the nanoscale'', project A 10 (P.H.) and Deutsche
Forschungsgemeinschaft project HA 1517/13-4 (P.H.).
\bibliographystyle{unsrt}

\begin{thebibliography}{10}
\bibitem{inversion} The explicit expression for the phase involving
  the $\arctan$ has to be understood in the sense of adding multiples
  of $\pi$ to make it a continuously growing function of time.
\bibitem{divf} The statement requires the function $f$ defined in 
   Eq.~(\ref{eq:bogmist1b}) to remain finite for $\phi^N=\pi/2 + n\pi,
   n\in N$; this, however, is no severe restriction.
\end{thebibliography}


\end{document}